%
%
%
%
\def\Teff{{\it T}\lower.5ex\hbox{\rm eff}}

\def\arcmin{$^\prime$}
\def\arcsec{$^{\prime\prime}$}
\def\hrs{\raise.5ex\hbox{h}}
\def\min{\raise.5ex\hbox{m}}
\def\sec{\raise.5ex\hbox{s}}
%
%
\def\lsim{<\kern-1.7ex $\lower0.85ex\hbox{$\sim$}$\ }     
\def\gsim{>\kern-1.7ex $\lower0.85ex\hbox{$\sim$}$\ }     
\def\element#1 #2{${}^{#1}$#2}
%
%
\def\MH{{M\lower.5ex\hbox{H}}}
\def\MHe{{M\lower.5ex\hbox{He}}}
\def\Mstar{{M\lower.5ex\hbox{$\star$}}}
\def\etal{{\it et al\/}.}
\def\ie{{\it i.e\/}.}

\def\cf{{\it cf}.}

%
%
\hyphenation{nucleo-cos-mo-chron-ology}
\hyphenation{Nucleo-cos-mo-chron-ology}
\hyphenation{chron-o-meter}

\def\today{\number\day\enspace
     \ifcase\month\or January\or February\or March\or April\or May\or
     June\or July\or August\or September\or October\or
     November\or December\fi \enspace\number\year}
\def\clock{\count0=\time \divide\count0 by 60
    \count1=\count0 \multiply\count1 by -60 \advance\count1 by \time
    \number\count0:\ifnum\count1<10{0\number\count1}\else\number\count1\fi}
\footline={\ifnum\pageno=1 \hss \hfill \hss
           \else \hss\rm -- \folio\ -- \hss\fi}

\overfullrule=0pt 
\raggedbottom
\def\push#1{\hbox to 1truein{\hfill #1}}

 
\def\arcsec{\ifmmode^{\prime\prime}\else $^{\prime\prime}$\fi}
\def\arcmin{\ifmmode^{\prime}\else $^{\prime}$\fi}
\def\deg{\ifmmode^\circ\else$^\circ$\fi}

\def\Teff{{\it T}\lower.5ex\hbox{\rm eff}}
\def\kms{\ifmmode{{\rm km~s}^{-1}}\else{km~s$^{-1}$}\fi}
\def\pervol{\ifmmode{{\rm cm}^{-3}}\else{cm$^{-3}$}\fi}
\def\perarea{\ifmmode{{\rm cm}^{-2}}\else{cm$^{-2}$}\fi}
\def\mum{\ifmmode{\mu {\rm m}}\else{$\mu$m}\fi}
\def\respwr{\ifmmode{\lambda /\Delta\lambda}
    \else{$\lambda /\Delta\lambda$}\fi}
\def\whz{\ifmmode{{\rm W~Hz}^{-1/2}}\else{W~Hz$^{-1/2}$}\fi}
\def\wig#1{\mathrel{\hbox{\hbox to 0pt{%
    \lower.5ex\hbox{$\sim$}\hss}\raise.4ex\hbox{$#1$}}}}
\def\lsim{\wig <}
\def\gsim{\wig >}
\def\sqr#1#2{{\vcenter{\vbox{\hrule height.#2pt
    \hbox{\vrule width.#2pt height#1pt \kern#1pt
    \vrule width.#2pt}
    \hrule height.#2pt}}}}

\def\boxit#1{\vbox{\hrule\hbox{\vrule\kern3pt
   \vbox{\kern3pt#1\kern3pt}\kern3pt\vrule}\hrule}}


\def\sec{{\rm sec}}

\def\etal{{\it et~al.\/}}

\def\cf{{\it cf.\/}}
\def\ie{{\it i.e.\/}}

\hyphenation{nucleo-cos-mo-chron-ology}  
\hyphenation{Nucleo-cos-mo-chron-ology}
\hyphenation{chron-o-meter}
\hyphenation{Ro-ches-ter}
\hyphenation{mol-e-cules}
 

\def\jref#1 #2 #3 #4 {{\par\noindent \hangindent=3em \hangafter=1
      \advance \rightskip by 5em #1, {\it#2}, {\bf#3}, #4.\par}}
\def\ref#1{{\par\noindent \hangindent=3em \hangafter=1
      \advance \rightskip by 5em #1\par}}
 
 
\newcount\eqnum
\def\nexteq{\global\advance\eqnum by1 \eqno(\number\eqnum)}
\def\lasteq#1{\if)#1[\number\eqnum]\else(\number\eqnum)\fi#1}
\def\preveq#1#2{{\advance\eqnum by-#1
    \if)#2[\number\eqnum]\else(\number\eqnum)\fi}#2}
 

\def\tableheight{\vrule width 0pt height 8.5pt depth 3.5pt}
{\catcode`|=\active \catcode`&=\active
    \gdef\tabledelim{\catcode`|=\active \let|=\vbar
                     \catcode`&=\active \let&=\nobar} }
\def\table{\begingroup
    \def\twidth{\hsize}
    \def\tablewidth##1{\def\twidth{##1}}
    \def\defaultheight{\vrule width 0pt height 8.5pt depth 3.5pt}
    \def\heightdepth##1{\dimen0=##1
        \ifdim\dimen0>5pt
            \divide\dimen0 by 2 \advance\dimen0 by 2.5pt
            \dimen1=\dimen0 \advance\dimen1 by -5pt
            \vrule width 0pt height \the\dimen0  depth \the\dimen1
        \else  \divide\dimen0 by 2
            \vrule width 0pt height \the\dimen0  depth \the\dimen0 \fi}
    \def\spacing##1{\def\defaultheight{\heightdepth{##1}}}
    \def\nextheight##1{\noalign{\gdef\tableheight{\heightdepth{##1}}}}
    \def\end{\cr\noalign{\gdef\tableheight{\defaultheight}}}
    \def\zerowidth##1{\omit\hidewidth ##1 \hidewidth}
    \def\hline{\noalign{\hrule}}
    \def\skip##1{\noalign{\vskip##1}}
    \def\bskip##1{\noalign{\hbox to \twidth{\vrule height##1 depth 0pt \hfil
        \vrule height##1 depth 0pt}}}
    \def\header##1{\noalign{\hbox to \twidth{\hfil ##1 \unskip\hfil}}}
    \def\bheader##1{\noalign{\hbox to \twidth{\vrule\hfil ##1
        \unskip\hfil\vrule}}}
    \def\spanloop{\span\omit \advance\mscount by -1}
    \def\extend##1##2{\omit
        \mscount=##1 \multiply\mscount by 2 \advance\mscount by -1
        \loop\ifnum\mscount>1 \spanloop\repeat \ \hfil ##2 \unskip\hfil}
    \def\vbar{&\vrule&}
    \def\nobar{&&}
    \def\hdash##1{ \noalign{ \relax \gdef\tableheight{\heightdepth{0pt}}
        \toks0={} \count0=1 \count1=0 \putout##1\end
        \toks0=\expandafter{\the\toks0 &\end} \xdef\piggy{\the\toks0} }
        \piggy}
    \let\e=\expandafter
    \def\putspace{\ifnum\count0>1 \advance\count0 by -1
        \toks0=\e\e\e{\the\e\toks0\e&\e\multispan\e{\the\count0}\hfill}
        \fi \count0=0 }
    \def\putrule{\ifnum\count1>0 \advance\count1 by 1
  \toks0=\e\e\e{\the\e\toks0\e&\e\multispan\e{\the\count1}\leaders\hrule\hfill}
        \fi \count1=0 }
    \def\putout##1{\ifx##1\end \putspace \putrule \let\next=\relax
        \else \let\next=\putout
            \ifx##1- \advance\count1 by 2 \putspace
            \else    \advance\count0 by 2 \putrule \fi \fi \next}   }
\def\tablespec#1{
    \def\vdimens{\noexpand\tableheight}
    \def\tabby{\tabskip=0pt plus100pt minus100pt}
    \def\r{&################\tabby&\hfil################\unskip}
    \def\c{&################\tabby&\hfil################\unskip\hfil}
    \def\l{&################\tabby&################\unskip\hfil}
    \edef\templ{\noexpand\vdimens ########\unskip  #1
         \unskip&########\tabskip=0pt&########\cr}
    \tabledelim
    \edef\body##1{ \vbox{
        \tabskip=0pt \offinterlineskip
        \halign to \twidth {\templ ##1}}}
    \edef\sbody##1{ {
        \tabskip=0pt \offinterlineskip
        \halign to \twidth {\templ ##1}}}
}
 

\def\ltextindent#1{\indent\llap{\hbox to \parindent {#1\hfil}}\ignorespaces}
 
\def\mathfont#1{{
    #1\count20=\fam\multiply\count20 by "100\advance\count20 by "7000
    \count21=`a \advance\count21 by - 1
    \count22=\count21\advance\count22 by \count20
    \loop \advance\count22 by 1 \advance\count21 by 1
             \global\mathcode\count21=\count22
    \ifnum \count21<`z \repeat
    \count21=`A \advance\count21 by - 1
    \count22=\count21\advance\count22 by \count20
    \loop \advance\count22 by 1 \advance\count21 by 1
             \global\mathcode\count21=\count22
    \ifnum \count21<`Z \repeat}}

\outer\def\section#1\par{\vskip 12pt plus 10pt minus 4pt
    \centerline{\bf#1}\nobreak\vskip 5pt plus 3pt minus 2pt}
 
 
\font\title=cmbx10 scaled \magstep2
 
\font\twelverm = cmr10 scaled\magstephalf \font\ninerm = cmr9
       \font\sevenrm = cmr7
\font\twelvei = cmmi10 scaled\magstephalf \font\ninei = cmmi9
       \font\seveni = cmmi7
\font\twelveit = cmti10 scaled\magstephalf 
       
\font\twelvesy = cmsy10 scaled\magstephalf \font\ninesy = cmsy9
       \font\sevensy = cmsy7
\font\twelvebf = cmbx10 scaled\magstephalf \font\ninebf = cmbx9
       \font\sevenbf = cmbx7
\font\twelvesl = cmsl10 scaled\magstephalf
\font\twelveit = cmti10 scaled\magstephalf
\font\twelvett = cmtt10 scaled\magstephalf
%
\textfont0 = \twelverm \scriptfont0 = \ninerm 
       \scriptscriptfont0 = \sevenrm
       \def\rm{\fam0 \twelverm}
\textfont1 = \twelvei \scriptfont1 = \ninei 
       \scriptscriptfont1 = \seveni
       
\textfont2 = \twelvesy \scriptfont2 = \ninesy 
       \scriptscriptfont2 = \sevensy
       
\newfam\itfam \def\it{\fam\itfam \twelveit} \textfont\itfam=\twelveit
\newfam\slfam  \textfont\slfam=\twelvesl
\newfam\bffam \def\bf{\fam\bffam \twelvebf} \textfont\bffam=\twelvebf
       \scriptfont\bffam=\ninebf \scriptscriptfont\bffam=\sevenbf
\newfam\ttfam  \textfont\ttfam=\twelvett
\rm
\hsize=6.5in 
\vsize=8.9in 
\baselineskip=23pt 
  \baselineskip=16pt 
\raggedbottom
%
\dimen1=\baselineskip \multiply\dimen1 by 3 \divide\dimen1 by 4
\dimen2=\dimen1 \divide\dimen2 by 2
%
%
\nopagenumbers
\headline={\ifnum\pageno=1 \hss\thinspace\hss 
     \else\hss\folio\hss \fi}
%
%
 
     \nobreak
%
%
\count10 = 0
\count11 = 0
\count12 = 0
\def\section#1{ 
    \vbox to \dimen1 {\vfill}
    \global\advance\count10 by 1
    \centerline{\uppercase\expandafter{\romannumeral\count10}.\ \bf {#1}}
    \global\count11=96  
    \global\count12=0   
    \nobreak
    \vskip \dimen1}
    \nobreak
%
%
\def\subsection#1{\global\advance\count11 by 1 
    \vskip \parskip  \vskip \dimen2
    \centerline{{\it {\char\number\count11}\/})\ {\it #1}} 
    \global\count12=0   
    \nobreak\vskip \dimen2\nobreak}
 
%
%
\def\subsubsection#1{\global\advance\count12 by 1 
    \goodbreak \vskip \parskip  \vskip \dimen2
    \centerline{\it {\expandafter{\romannumeral\count12}})\ {\it #1}} 
    \nobreak\vskip \dimen2\nobreak}
%
%
\def\boxit#1{\vbox{\hrule\hbox{\vrule\kern3pt
      \vbox{\kern3pt#1\kern3pt}\kern3pt\vrule}\hrule}}
%
%
\def\refindent{\advance\leftskip by 24pt \parindent=-24pt}

%

%

%

%

%

%

\magnification=1000
\hsize=6.0truein
\vsize=8.5truein
\hoffset=0.0truein
      
\ \vskip 0.5truein
\centerline{\title X-ray Spectral Characteristics of GINGA Gamma-Ray Bursts}
\vskip 3pt
\centerline{T. E. Strohmayer\footnote{$^1$}{Current address: GSFC,
stroh@pcasrv.gsfc.nasa.gov}, E. E. Fenimore
\footnote{$^2$}{efenimore@lanl.gov}}
\centerline{Los Alamos National Laboratory}
\centerline{Los Alamos, New Mexico 87545}
\vskip 2pt
\centerline{T. Murakami}
\centerline{Institute of Space and Astronautical Science}
\centerline{1-1 Yoshinodai 3, Sagamihara, Kanagawa, 229 Japan}
\vskip 2pt
\centerline{A. Yoshida}
\centerline{The Institute of Physical and Chemical Research (RIKEN)}
\centerline{2-1, Hirosawa, Wako, Saitama, 351-01, Japan}
\vskip 2pt
\vskip 5pt
\centerline{Received: \_\_\_\_\_\_\_\_\_  Accepted: \_\_\_\_\_\_\_\_\_\_}

\vskip 5pt
\centerline{\bf Abstract}
\vskip 5pt

We have investigated the spectral characteristics of a sample of bright 
$\gamma$-ray bursts detected with the $\gamma$-ray burst sensors aboard the 
satellite {\it Ginga}. This instrument employed a 
proportional and scintillation
counter to provide sensitivity to photons in the 2 - 400 keV
region and as such provided a unique opportunity to characterize the 
largely unexplored X-ray properties of $\gamma$-ray bursts. The photon spectra 
of the {\it Ginga} bursts are well described by
a low energy slope, a bend energy, and a high energy slope.
 In the energy range where
they can be compared, this result 
is consistent with burst spectral analyses obtained from the BATSE 
experiment aboard the {\it Compton Observatory}. However, below 20 keV we find 
evidence for a positive spectral number index  in approximately
40\% of our burst sample, with some evidence for a strong rolloff
at lower energies in a 
few events.
There is a correlation (Pearson's $r$ = -0.62) between the low energy
slope and the bend energy.
We find that the distribution of spectral bend energies
extends below 10 keV.
There has been some concern in cosmological models of GRBs that the bend
energy covers only a small dynamic range.  Our result extends the observed
dynamic range and, since we observe bend energies down
to the limit of our instrument,
 perhaps observations have  not yet limited the range.
The {\it Ginga} trigger range was virtually the same
as BATSE's,
yet we find a different range of fit parameters.
One possible explanation might be that GRBs have two break energies, one
often in the 50 to 500 keV range and the other near 5 keV.  Both BATSE and
{\it Ginga} fit with only a single break energy so BATSE tends to find breaks
near the center of its energy range and we tend to find
breaks in our energy range.
The observed ratio of energy emitted in the x-rays relative to the gamma-rays
can be much larger than
 a few percent and, in fact, is sometimes larger than
 unity.  The average for our 22 bursts is 24\%.
 We also investigated spectral evolution in two bursts.
 In these events we find strong
evidence for spectral softening as well as a correlation between photon 
intensity and spectral hardness. We also find that the X-ray signal below 30 
keV itself softens in both of these events. There is one example of
a strong x-ray excess at low energy.  In addition to providing further 
constraints on $\gamma$-ray burst models, the description provided here of 
burst spectra down to 2 keV should prove useful to future planned efforts to 
detect bursts at X-ray energies.
\hfill\break
{\it Subject Headings:} Gamma Rays: Bursts - X-Rays: General
\hfill\break

\vfill\eject

\vskip 10pt
\noindent{\bf 1.  Introduction}
\vskip 10pt

Twenty-five years after their discovery, $\gamma$-ray bursts continue to defy
explanation. 
Analysis of burst energy spectra remains one of the principal methods for
determining the physical processes responsible for these events. 
Recent analysis of spectroscopy data from 
the Burst and Transient Source Experiment (BATSE) on 
the {\it Compton Gamma Ray 
Observatory} (CGRO) has demonstrated the diversity of burst
spectral continua in 
the $\approx 30 - 3000$ keV range (\cf\ Band \etal\ 1993; hereafter B93). 
However, few instruments to date have probed the X-ray regime of $\gamma$-ray 
bursts between 2 and 20 keV. 

The first detections of X-rays in the 1-8 keV range from $\gamma$-ray bursts
were reported by Wheaton \etal\ (1973), Trombka \etal\ (1974), and Metzger 
\etal\ (1974). Later results from the P78-1 (Laros \etal\ 1984) and {\it 
Hakucho} (Katoh \etal\ 1984) satellites confirmed 
that $\gamma$-ray bursts often
produce significant X-ray emission. In addition, these satellites provided the 
first evidence suggesting that the X-ray emission might outlast the main 
$\gamma$-ray event in some bursts (so called X-ray tails). These early data 
provided only modest spectral resolution in the X-ray regime, 
however, based on 
their results, the $\gamma$-ray burst 
detector (GBD) flown aboard the {\it Ginga} satellite 
was specifically designed 
to investigate burst spectra in the X-ray regime (Murakami \etal\ 1989). 
{\it Ginga} was launched in 
February of 1987, and the GBD was operational from March, 1987 until the 
reentry of the spacecraft in October, 1991. Several important results 
have emerged from the study of burst spectra recorded 
with the {\it Ginga} GBD. Harmonically spaced line features have been 
reported from several {\it Ginga} events (Murakami \etal\ 1988; 
Fenimore \etal\ 
1988; Yoshida \etal\ 1992; Graziani \etal\ 1992), and have been interpreted as
due to cyclotron absorption in the strong magnetic field of a neutron star.
Recently, preliminary results indicate that BATSE has also detected
absorption lines
(Briggs et al, 1997).   
In 
addition to line features, {\it Ginga} has observed X-ray tails in a number of 
bursts, as well as x-ray preactivity in one event (Yoshida \etal\ 1989;
Murakami \etal\ 1991; \& Murakami \etal\ 1992b).
More recently, the BeppoSax satellite has observed several bursts in both
x-rays and gamma-rays (Piro, et al. 1997) and discovered soft x-ray
afterglows (Costa et al., 1997).  
This latter discovery has opened the way for the
long sought GRB counterparts, including one with a measured redshift
(Metzger et al., 1997).
 Here, we provide for 22 GRBs
a compilation and description of $\gamma$-ray burst spectral continua
emphasizing the X-ray spectral characteristics.

Before BATSE, the analysis of $\gamma$-ray burst 
spectra suggested that the continua can be
fitted adequately by a range of models, with power law, optically thin thermal 
bremsstrahlung and thermal synchrotron formulae providing acceptable fits to 
many spectra (\cf\ Mazets \etal\ 1982; Matz \etal\ 1985; and Hurley 1989). 
Although these functional forms adequately describe 
many burst spectra it is not
at all clear that the corresponding physical processes are responsible for the 
burst spectra, because spectral shapes 
inconsistent with each other may often be
consistent with the same data. 
The compilation by B93 of average spectra
from a large sample of BATSE bursts provides the most detailed statistical 
description presently available of burst continua in the range from $\approx 
30 - 3000$ keV.
(Also see Schaefer \etal\ 1994.)
 It is our purpose here to provide
an extension of this statistical description of burst spectra 
down to 2 keV using an analysis of bright bursts observed by 
the {\it Ginga} GBD. In addition to providing further 
constraints on proposed burst source models it is our hope 
that the information provided here will be useful 
to future soft X-ray experiments which will require 
the range of GRB spectral characteristics
to estimate,
for example, the expected rate of burst detections. 
To facilitate comparison of our results with the published BATSE spectra 
we adopt the same burst spectral form as described in B93. 
Although it does not imply any particular physical process,
this form provides an adequate description of BATSE bursts and the
{\it Ginga} bursts studied here.
Given the excellent statistics generated by the BATSE detectors, the fact
that the B93 shape can adequately fit the GRB spectrum is a strong
argument that it should be used even though it does not imply any
physical process.

We begin in \S 2 with a brief description of the GBD instruments, as well
as a discussion of the analysis techniques employed, including a discussion of 
the constraints placed on us by the lack of positional information for many of 
the {\it Ginga} bursts. In \S 3 we describe the spectral properties of the 
{\it Ginga} sample and compare these results with the previously reported 
BATSE spectra. In \S 4 we investigate the spectral evolution of a smaller 
number of bright bursts.
In \S 5 we summarize the x-ray emission relative to the gamma-ray emission.
We conclude in \S 6 with a summary and discussion of
our principal results.

\vskip 10pt
\noindent{\bf 2. Instrumental Summary and Data Analysis Techniques}
\vskip 10pt

The GBD aboard {\it Ginga} consisted of a proportional 
counter (PC) sensitive to
photons in the 2-25 keV range and a scintillation counter (SC) recording 
photons with energies between 15-400 keV. Each detector had an $\approx$ 60 
cm$^2$ effective area. The PC and SC provided 16 and 32 channel spectra, 
respectively, over their indicated energy ranges. The detectors were 
uncollimated except for the presence of shielding to reduce backside 
illumination and the mechanical support for the PC window.
The field of view was effectively  $\pi$ steradian for both
the PC and the SC. In
burst mode the GBD recorded spectral data from the 
PC and SC at 0.5 s intervals 
for 16 seconds prior to the burst trigger time, and for 48 seconds after the 
trigger. This Memory Read Out (MRO) data was used for 
many of the spectral fits 
described here. In the event that MRO data was not available for a burst, we
utilized the spectral data from the ``real time'' telemetry modes.
 For these bursts, spectral data were
available with either 2, 16, or 64 second accumulations.  For the longer
accumulations, spectral 
studies were not generally feasible.
 See Murakami \etal\ (1989) 
for more information concerning the detectors.

Before a model can be fitted to the data the background spectrum must be 
estimated and subtracted from the counts. 
For a majority of the analyzed bursts a linear fit to the MRO data in 
each energy channel provided a reasonable fit to the pre- and post-burst data.
In a few cases, a quadratic fit was used.  Events with large variations
in the background were rejected from this analysis.
 For several events, the background was estimated 
from real-time data which was generally available for a longer 
time period both 
before and after the event. The background subtracted spectra were then fitted 
using a standard $\chi^2$ minimization technique. 
In this method the input model
spectrum is first folded through the detector response functions, thereby 
providing count predictions for each energy channel, 
these predictions are then 
compared to the observed counts using a $\chi^2$ statistic.
The model parameters 
are adjusted iteratively until a minimum in $\chi^2$ is obtained. As
mentioned previously, we have adopted the spectral 
model employed by B93 because of its relative simplicity and ability to 
accurately characterize a wide range of spectral continua, in addition this 
choice facilitates direct comparison of our results with those 
obtained from the
BATSE bursts. This model has the form
$$
N(E) = A \left ( {E \over 100 {\rm keV}}\right )^{\alpha} \exp (-E/E_0) \;\;\; 
, \;\;\; (\alpha -\beta)E_0 \ge E
$$
$$
N(E) = A\left [ {(\alpha -\beta)E_0 \over 100 {\rm keV} } \right ] ^{\alpha 
-\beta} \exp (\beta -\alpha) \left ( {E\over 100 {\rm keV}} \right )^{\beta}
\;\;\; , \;\;\; (\alpha -\beta)E_0 \le E , \eqno(1)
$$
where $A$ is an overall scale factor, $\alpha$ is the 
low energy spectral index,
$\beta$ is the high energy spectral index, and $E_0$ is the exponential cutoff 
or bend energy. In addition, to facilitate comparison with previous 
calculations we have restricted the ranges of $\beta$ to be greater than -5.0.
Recently, this spectral form has been characterized with the peak of
the $\nu F_{\nu}$ distribution (i.e., $E_p$).  The conversion is that
$E_p = (\alpha+2)E_0$.

The {\it Ginga} GBD was in operation from March 1987 to October 1991. During 
this time $\approx 120$ $\gamma$-ray bursts were identified (\cf\ Ogasaka 
\etal\ 1991; \& Fenimore \etal\ 1993). We selected from
this group a sample of 22 events for which good 
spectral data were available and
for which we could be reasonably certain that the burst occurred within the 
forward, $\pi$ steradian field of view of the detectors (front-side events). 
Front-side events are easily identified because they have consistent fluxes
 in the
energy range observed by both instruments (15 to 25 keV).
This sample includes 18 of the events searched for spectral lines (Fenimore 
\etal\ 1993).
Excluded bursts usually show strong rolloffs below 25 keV
and spectral fits with an absorption component due to aluminum 
(a principal spacecraft material) are consistent with this interpretation.
Sky positions for four of the events in our 
sample are known because of simultaneous detections with 
either BATSE or WATCH. 
Incidence angles into GBD for bursts 910429, 910717, and 910814
(9.3, 34.3, 37.0 degrees, respectively)
 were derived
from knowledge of the GBD orientation at the time of the burst and the 
published BATSE positions (Fishman \etal\ 1994). Burst 900126 was seen by the 
WATCH experiment and the incidence angle (50 degrees) was
derived using the published
position (Lund 1992).  

For the remaining events, the incidence angle $\theta$ of the photons into the 
detectors is uncertain ($0^{\circ}$ equals normal incidence). Since the 
detector response is a function of this angle, the inferred 
source spectrum and 
thus peak intensity of these events is also somewhat uncertain.
We selected 37 degrees for the incidence angle when the angle was unknown.
This is a typical angle considering that the mechanical support
for the window
on the PC acts as
a collimator limiting the field of view to
an opening angle of
$\sim \pm 60$ degrees.  We have used
Monte Carlo simulations to evaluate the impact of the unknown incidence
angle.   For each simulation, we selected a random incidence angle between
0 and 60 degrees.  We calculated a response matrix and used it with the burst's
best fit parameters in Equation (1) to generate simulated spectra.  Background
was added and Poisson statistics applied.  We then analyzed the simulated
data the same way we analyzed the burst data: an estimated background
was subtracted and the best fit parameters for Equation (1) were
found based on a
response matrix corresponding to 37 degrees. 

The uncertain incidence angle
usually only affects $\alpha$ and $E_0$ in Equation (1) so we will emphasize
the affect of the uncertain angle on those parameters.  Figure 1 shows
typical examples covering interesting values of $\alpha$ (i.e., positive 
$\alpha$'s: 0.74, 0.22, and 1.67 for GB900901, GB870303, and GB900322,
respectively).  Each cross represents the best fit $\alpha$ and $E_0$ from
a simulation. We show 100 simulations per burst for clarity although we did 
$10^3$ simulations.  The contour includes 68\% of the cases (based on
the $10^3$ simulations).  As was found by B93, the observations tend to
agree with a range of solutions satisfying $\alpha \propto \log E_0$.
Figure 1 shows the range of uncertainty due to both the unknown incidence
angle and counting statistics.

\vskip 10pt
\noindent{\bf 3. X-ray Spectral Characteristics of GINGA Bursts}
\vskip 10pt

For each burst in the sample we selected a time interval $\Delta T$ over which 
to sum the counts in each spectral channel. This selection was based on the 
need to increase the signal to noise in the
computed spectrum and the desire to
constrain the spectrum near the peak of the time profile. 
Sometimes, $\Delta T$ is limited by the available time resolution in
the real time data.
If the burst has a smaller duration than our resolution, we will underestimate
its intensity (e.g., GB910814).
Spectra can evolve 
substantially during the course of an event (\cf\ Mazets \etal\ 1981;
Golenetskii \etal\ 1983;
 Norris
\etal\ 1986; Band \etal\ 1992; Kargatis \etal\ 1994; 
Ford \etal\ 1995; and \S 4). We attempted to compute a spectrum consistent 
with the most intense portion of each burst's time profile. 
For real-time events we endeavored to compute the 
spectra from the time ranges which encompassed the majority of the burst. We 
employed all spectral channels in most of the model fits except for the lowest 
PC channel which has extremely small effective area, 
for a total of 47 channels.
To illustrate these spectral fits we show in Figure 2
the best-fit spectrum and indicate the contribution of $\chi^2$ for each
 channel for each
 burst analyzed.
The burst nomenclature
specifies the year/month/day on which each event was detected. 
In the event that
more than one burst occurred on a given day, a letter is assigned to the burst
according to the order of occurrence. 

We choose the format of Figure 2 to simultaneously show the best-fit
spectra
as well as how well each spectral channel fits the observations.  Note
that the ordinate is spectral number flux, not counts.
Let $O_i$ be the observed net count rate for the $i$-th channel and
$M_i$ be the predicted net rate from the model, $N(E)$.  Let
$\sigma_i^2$ be the variance on $O_i$.  The solid line in these
figures is $N(E)$ and each channel is represented by a point plotted
at $O_iN(E)/M_i$ with a vertical line length of $\pm \sigma_i N(E)/M_i$.
  Thus, for each point,  the distance from the point to
the line divided by the length of the error bar is $(M_i -
O_i)/\sigma_i = \chi_i$, that is, it  gives the contribution of the point to
the total $\chi^2$.  The alternative ways to present these data would be
to plot the residuals $M_i - O_i\pm\sigma_i$  or
to plot $M_i$ and $O_i\pm\sigma_i$.
Each method emphasizes a different aspect of the fitting: the model
photon spectrum by the first, the residuals by the second, and the observed
counts by the third.
It is equally possible to visualize
the contribution of each channel
to $\chi^2$ in all three presentations.  All three methods are only
as valid as the assumed $N(E)$.
The third method is often characterized as being independent of the model
but, of course, the only useful information is whether the observed counts
differ significantly from the predicted counts, and those predicted
counts depend on the model.
 One most be careful in all three
presentations to avoid interpreting deviations of $O_i$ from $M_i$ as
implying that the spectrum might be locally different from the assumed
$N(E)$ because is all three cases, $M_i$ depends on a wide
range (i.e., from $E$ to $\infty$) of $N(E)$ so is very model dependent.
The temptation to view
local differences between $O_i$ and $M_i$ as evidence that $N(E)$
should be different is probably strongest for  our format since
$N(E)$ is present in the figure.
However, all three formats are equivalent on this score:
the quality of
a model fit
should be judged solely on the basis of the total $\chi^2$ value
and not on how any one point (or group of points) matches
(or mismatches) the best-fit
model. 
Our format is misleading if one attempts to interpret the
deviations as implying that the best-fit model should locally change
to ``agree'' with the observations.  That is why this format is
inappropriate for presentations of absorption line candidates or to
argue that a model needs to be changed. However, in this work, we are
showing {\it acceptable} fits and
we will not argue for changes in the best fit model.
This format shows the best fit spectral function and how well each point
agrees with it.  As such, it is a good method for visualizing the low energy
behavior of GRBs.

The results of the spectral fits for each burst are summarized in Table 1. 
Here, $\Delta T$ specifies the time interval for calculation of the 
spectrum,
$R_{x/g}$ is the ratio of the emitted energy in x-rays relative to
the gamma-rays (see \S 5),
$E_0$, $\alpha$, and $\beta$ are the best-fit model parameters, and
$\chi^2_r$ is the reduced $\chi^2$ for the fit. 
Each spectral fit was performed 
with 43 degrees of freedom, and the quoted uncertainties 
for $\beta$
are $1\sigma$ estimates
assuming one parameter of interest.
The strong coupling in the B93 spectral shape between $\alpha$ and $E_0$
means that one should not quote separate confidence regions for $\alpha$
and $E_0$ (see below).

Previous spectral analyses have been reported for some of the bursts listed 
here, mostly with regard to the presence of lines in their spectra, but also 
because of interesting X-ray properties (\cf\ Murakami \etal\ 1988; Fenimore 
\etal\ 1988; Wang \etal\ 1989; Yoshida \etal\ 1992; Yoshida \etal 1989; \& 
Murakami \etal\ 1991), here we have restricted our interest to the spectral 
continua of the bursts. In Table 1 we also give references
to previous analyses of these bursts (last column of Table 1). 

The spectral fits are generally acceptable, with $\chi^2_r$ of order unity for 
most of the bursts. In agreement with B93 we find that a range in these model 
parameter values is required to adequately 
describe $\gamma$-ray burst spectra. 
In Figure 3 we show a graphical comparison between
the BATSE results taken from 
B93 and our results for the {\it Ginga} sample. 
Figure 3a contains the best-fit
spectra for the 22 bursts in the {\it Ginga} sample, while Figure 3b displays
the 54 spectra computed from the BATSE data of B93. In each case the
spectra are plotted only over the nominal band-pass of each instrument (2-400 
keV and 20-3000 keV for {\it Ginga} and BATSE respectively). 
For BATSE, bursts were measured over slightly different bandpasses due to 
different detectors having different gains.  See B93 for the actual measured
range of each burst. The 
spectra have been normalized to 1.0 photon/keV at 100 keV. Between 20 and 400 
keV the two samples are in substantial agreement, 
and span a consistent range of
spectral shapes. 

Of particular interest is the behavior of the {\it Ginga} 
sample at X-ray energies. About 40\%  of the bursts in the sample show a 
positive spectral number index below 20 keV (i.e., $\alpha >
0$),
 with the suggestion of 
 rolloff toward lower energies in a few of the bursts ($\alpha$ as
large as $\sim +1.5$).
Unfortunately, the 
lack of data below 1 keV, and the often weak signal 
below 5-10 keV precludes us 
from establishing the physical process (photoelectric absorption,
self-absorption) that may be 
involved in specific bursts. 
The remainder of the burst spectra continue to increase below 10 keV.
Observations of the low-energy asymptote can place serious constraints
on several GRB models, most notably the synchrotron shock model which
predicts that $\alpha$ should be between -3/2 and -1/2 (Katz 1994).
Crider et al. (1997) uses examples from BATSE to argue that some GRB
violate these limits during some {\it time-resolved} samples.  We find
violations of these limit in the {\it time-integrated} events.

Figure  4 summarizes the low energy behavior of GRBs as a scatter plot
of the bursts $\alpha$'s and $E_0$'s.  In Figure 4a, we show the 68\%
confidence regions for 20 of the bursts in Table 1.  These confidence
regions were calculated the same way as   in Figure 1.  For two bursts
we only show the best fit parameters as solid squares because a simple
power law could nearly fit the entire spectra.  Note for GB881009 that
$\alpha$ is nearly equal to $\beta$ and for GB880205, the bend energy
is well above our energy range.  Although we give the formal best fit
parameters in Table 1, effectively, $E_0$ was undetermined or not
necessary for the fit in those two bursts.  We repeated the analysis
of Figure 4a in Figure 4b except we did each simulation at the angle
for which we analyzed the burst (37 degrees, in most cases). Thus,
Figure 4b indicates the size of the confidence regions if we knew the
incidence angle.  By comparing Figure 4a and Figure 4b, we see that
the lack of knowledge of the incidence angle into {\it Ginga} does not
introduce much more uncertainty than the counting statistics.
On average, because of the uncertain incidence angle, the confidence region
for $E_0$ is 22\% larger and the confidence region for $\alpha$ is larger
by 0.06.

In Figure 4c we combine the results from this paper and B93.
  The open squares
are the BATSE results from B93 and the solid squares are the {\it Ginga}
results presented here. 
Many of the {\it Ginga} points lie within the range found by BATSE.
However, the lowest $E_0$ found by BATSE was 14 keV (set, of course, by the
lowest energy observed by BATSE).  Ginga extends $E_0$ values down to 2
keV.  BATSE had a small fraction (15\%) of events with $\alpha > 0$
whereas {\it Ginga} has 40\% of events with $\alpha > 0$.
 In general this is because there is a correlation between $\alpha$ and
$E_0$ such that the lower energy range of {\it Ginga} samples a parameter
space with more events with $\alpha > 0$.  For the 76 points, the Pearson's
$r$ coefficient is -0.62.
The formal significance is about 4$\sigma$ although that
ignores the complicated error bars that
are caused by the fact that the observations tend to agree with a range
of $\alpha-E_0$. However, the existence of a correlation seems
reasonable: there are virtually no
events seen by BATSE at large $\alpha$, large $E_0$ and few low $\alpha$,
low $E_0$ events seen by {\it Ginga}.
  The error bars (Fig. 4a) are large on the {\it Ginga} points, but not  large
enough to indicate that they all should fall where the average BATSE events
occurred.
We note that GB900126 is an event for which we know the incidence angle.
This event is important to our conclusions because it has one of the
largest $\alpha$'s and there is no uncertainty due to an unknown
incidence angle.
In Figure 4c we  show a representative confidence regions for
 one BATSE event from B93, burst 451.
We note that the confidence regions quoted in Table 4 of B93 were found
separately for $\alpha$ and $E_0$, ignoring the coupling, 
so are much smaller than the real
confidence regions and cannot be compared to ours.

In our sample of bursts, we find that $\alpha$ can be both less than
zero and greater than zero.  Negative $\alpha$'s are often seen in
time-integrated BATSE spectra.  Positive $\alpha$'s where the spectrum
rolls over at low energies are usually only seen in {\it time
resolved} BATSE spectra (Crider et al.
 1997).
 
The {\it Ginga} trigger range (50 to 400 keV) was virtually the same
as BATSE's. Thus, we do not think we are sampling a different population of
bursts, yet we get a different range of fit parameters.
The lack of events with $E_0$'s between 6 and 20 keV cannot
 be used to support two populations because we
do not have enough events.
One possible explanation might be that GRBs have two break energies, one
often in the 50 to 500 keV range and the other near 5 keV.  Both BATSE and
{\it Ginga} fit with only a single break energy so BATSE tends to find breaks
near the center of its energy range and we tend to find
breaks in our energy range.
Without good high energy observations of bursts with low $E_0$, it is
difficult to know whether they also have a high energy bend.

Preece et al (1996) utilized a low energy discriminator channel and detected
emission in excess of what would be expected from a fit at higher energy.
Preece et al (1996) report excesses in 15\% of the investigated BATSE bursts.
One of our bursts, GB880205, shows a clear strong excess at low energy
and two other {\it Ginga} bursts, GB880830 and GB910418,
 probably also shows an excess.
For GR910418, Preece et al (1996) also reported an excess.  From
the Preece et al (1996) result, we would expect about 3 of our bursts
to show an excess so we are consistent with the BATSE result.

\vskip 10pt
\noindent{\bf 4. Spectral Evolution of GINGA Bursts}
\vskip 10pt

It was recognized shortly after the discovery of the $\gamma$-ray bursts that 
their spectra show significant variation within a single event (Wheaton \etal\ 
1973; Metzger \etal\ 1974; Teegarden \& Cline 1980; \& Yoshida \etal\ 1989). 
Analysis of
{\it Solar Maximum Mission}, and BATSE data indicate that burst spectra 
tend to evolve from hard to soft, both for individual peaks in the temporal 
profile, as well as over the entire course of the burst, 
though counter-examples
to this general trend can be found (\cf\ Norris \etal\ 1986; and Band \etal\ 
1992, Band, 1997, Crider et al. 1997).
Previous analyses of the {\it Ginga} data identified several bursts with soft
X-ray preactivity as well as X-ray tails which generally showed softer spectra 
than the most intense portion of the event (Murakami \etal\ 1991; and Murakami 
\etal\ 1992b). 

Several of the {\it Ginga} bursts which we have analyzed were of sufficient 
duration and intensity to permit spectral evolution studies. Here we summarize 
our results for two events, bursts 901001 and 890929.
Incidence angles into the detectors for these events are not 
known, however, when comparing spectra computed from different time intervals 
within the same burst, the choice of incidence angle is not important,
since this 
angle did not vary during the event. We therefore used an angle of 
$37^{\circ}$ for these calculations as well. For each burst we subdivided the 
time profiles into several temporal regions and then computed the best-fit 
spectrum for each region. We used the same 
GRB model (eq. [1]) as in the previous calculations.

In Figure 5a we show the time profiles for burst
901001 in five different energy
ranges. The dotted vertical lines delineate the six temporal regions in which 
spectra were computed. Also shown is a linear estimate of the background 
computed from the pre- and post-burst signal. Note that the low energy, X-ray 
emission persists longer than the high energy emission, suggestive of overall 
softening of the burst spectrum (\cf\ Yoshida \etal\ 1992). 
In Figure 5b we show
the best-fit spectra 
computed for each indicated time period. 
We have again normalized each spectrum 
to 1.0 at 100 keV. The Figure legend identifies each spectrum with the time
interval for which it was computed. For completeness, the best-fit model 
parameters computed from each time interval are also given in Table 2. 
Comparison of the spectra computed from the first two time intervals suggests 
possible initial hardening during the rise to 
peak intensity. After the peak there is strong evidence for softening of the 
spectrum. In fact, for intervals $T_5$ and $T_6$ there is still a significant 
signal in the PC, however, there is no evidence of corresponding emission above
100 keV in the SC. 
To further investigate the relationship between intensity and
hardness we have
 computed  $E_p$ and
flux $F$ (between 2 and 400 keV) for each
time interval. The resulting values are plotted in the inset panel of Figure 
5b. The $1\sigma$ uncertainties were estimated from Monte Carlo simulations
using the best-fit spectral parameters derived from each time interval.
The direction of temporal evolution is
along the solid line, from upper right to lower left.
 Thus, there is
initial hardening during the rise to peak of the burst, followed by
softening for the remainder of the event. 
In Figures 6a, 6b and Table 2 we show
the corresponding results for burst 890929. 
There is also evidence for an X-ray 
tail in this event, and the softening of the spectrum 
during the course of this 
burst is also quite apparent, though no initial hardening during the rise to 
peak is evident.
 For these events
the X-ray signal also softens in addition to the
higher energy $\gamma$-rays.  

The hardness-intensity plots for these two bursts 
suggest a correlation between 
hardness and intensity similar to that reported by
Golenetskii \etal\ (1983) ,
Kargatis \etal\ (1994), and Bhat \etal\ (1994).
Linear fits to the $\log E_p$ versus $\log F$ data (\ie\ $\log E_p =
a\log F + b$) for bursts 901001 and 890929 give $a\approx 0.3$.
The results of our linear fits are shown in Table 3.

\vskip 10pt
\noindent{\bf 5. X-ray Emission Relative to Gamma-ray Emission}
\vskip 10pt

Early measurements of the x-rays associated with gamma-rays were 
fortuitous observations by collimated x-ray detectors that just happen
to catch a GRB in their field of view.  From a few events it appeared
that the amount of energy in x-rays was only a few percent (Laros et
al. 1984) confirming that GRBs were, indeed, a gamma-ray phenomena. 
  The {\it Ginga} experiment was designed with a wide field
of view to detect a sufficient number of events to determine the range
of x-ray characteristics.  Early reports from {\it Ginga} events
indicated that sometimes a much larger fraction of the emitted energy
was contained in the x-rays. For example, by comparing the signal in
the proportional counter (roughly 2 to 25 kev) to that of the
scintillator (roughly 15 to 400 kev), we reported an x-ray to
gamma-ray emission ratio up to $\sim $46\% (Yoshida \etal\ 1989).
  Such a ratio depends on the
bandpass for which it is evaluated.  One can use the parameters from
Table 1 in Equation (1) to determine the ratio of emission for
arbitrary bandpasses.  This would be useful, for example, to predict
the range of  emission that might be seen in instruments such as
ROSAT.  Here, we find the ratio of emission for a typical x-ray
bandpass (2 to 10 keV) compared to the BATSE energy range (50 to 300
keV).
The ratio is defined to be
$R_{x/g} = \int_{2}^{10}EN(E)dE/\int_{50}^{300}EN(E)dE$.
This ratio is listed in the third column of Table 1.  Figure 7
presents the distribution for the events analyzed in this paper.
Although the ratio is often a few percent, for some events the ratio
is near (or larger than) unity.
 Some GRBs actually have more energy
in the x-ray bandpass than the gamma-ray bandpass. 
The simply average of the 22 values is 24\%.
This large value arises because of the few events with nearly equal energy
in the x-ray and gamma-ray bandpass.  However, even the logarithmic average
is 7\%.

\vskip 10pt
\noindent{\bf 6. Discussion and Summary}
\vskip 10pt

Several of the results reported here are of particular interest.  
There is some concern in cosmological models of GRBs that the bend energy 
occurs only over a small range of values implying that the bulk Lorentz
factor only varies a little from burst to burst
 (Brainerd, 1994, 1997).
The  {\it Ginga}
bursts extends the range of observed bend energies from above 500 keV (with 
BATSE) to $\sim$ 3 keV ({\it Ginga}). This {\it Ginga} result implies that,
perhaps, we have not yet observed the lower limit on bend energies and,
thus, the range of dynamic range of bulk Lorentz factors has not been
limited.
 Regardless, any successful model for the burst sources must be 
able to explain the range of bend energies. 

The spectral evolution observed in bursts 901001 and 890929  
support the consensus that hard to soft spectral evolution is a 
prevalent feature of the burst mechanism.
 Both {\it Ginga} bursts analysed here would seem to 
fit this scenario. Initial soft to hard evolution has been previously
reported for burst 900126 (Murakami \etal\ 1991), and is also suggested by our 
analysis of burst 901001. Perhaps the timescale for the 
energization process is 
rapid enough in some bursts to preclude observation of the 
initial hardening of 
the spectrum.

Several future space missions have been planned with the goal of detecting 
bursts at soft X-ray energies.
 For example, the X-ray spectral
distribution of $\gamma$-ray bursts has consequences for the High Energy 
Transient Experiment (HETE) mission (\cf\ Ricker \etal\ 1992) which has as one 
of its main goals the identification of burst counterparts 
derived from localization of bursts using X-ray  observations.
The initial localization 
of sources by HETE depends on the detection of bursts with a coded aperture 
X-ray instrument operating from 2 - 25 keV.
Estimates of the number of 
bursts expected to be seen per year
can be made based on the range of x-ray spectra presented here.

We have investigated the X-ray spectral characteristics of a sample of 22
bright $\gamma$-ray bursts detected by the GBD which flew aboard {\it Ginga}. 
In the energy band where they can be compared, the range of spectral shapes 
which describe the {\it Ginga} sample are
consistent with those derived from
 analyses of bright BATSE bursts.
Consistent with Preece \etal\ 1996, we find at least one event (GB880205) that
has an x-ray excess relative to the GRB spectra from Equation 1.
 Moreover, the {\it Ginga} spectra
extend down to $\approx 2$ keV, and thus provide a useful characterization of 
burst spectra over an energy range at which burst spectra have not often been 
measured. In addition, we have investigated spectral evolution in two longer 
duration events. We found evidence for spectral softening during both of these 
events, along with a correlation between hardness and intensity.

This work was carried out under the auspices of the US Department of Energy.
It is a pleasure to acknowledge helpful discussions with Jean in 't Zand and 
Richard Epstein, and Tony Crider.
We also thank David Band for extensive discussions on the error bars for
$\alpha$ and $E_0$ as well as many comments on the manuscript.
Tony Crider kindly helped with the production of Figures 5 and 6.

\vfill\eject 

\vskip 10pt
\noindent{\bf References}
\vskip 10pt

\def\BB#1\par{\noindent\hangindent=20pt { #1 } \par \medskip }

\BB
Band, D., 1997, ApJ, in press

\BB Band, D. \etal\ 1992, in AIP Conference Proceedings 265, Gamma-Ray 
Bursts, ed. W. S. Paciesas \& G. J. Fishman (New York: AIP), 169.

\BB Band, D., \etal\ 1993, ApJ, 413, 281.



\BB Bhat, P. N., \etal\ 1994, ApJ, 426, 604.

\BB Brainerd, J. J., 1994, Ap. J. 428, 21

\BB Brainerd, J. J., 1997, Ap. J. submitted.

\BB Briggs, et al. 1997, talks given at the Aspen and Huntsville 1997
conferences on GRBs.

\BB Costa, E., et al., 1997, Nature, 387, 783

\BB
Crider, A., et al. 1997, ApJ, 479, L93

\BB Fenimore, E. E, \etal\ 1988, ApJ, 335, L71.


\BB Fenimore, E. E., \etal\ 1993, in AIP Conference Proceeding 280,
Compton Gamma-Ray Observatory, ed. M. Friedlander, N. Gehrels, \& D. Macomb, 
(New York:AIP), 917.


\BB Fishman, G., \etal\ 1994, ApJS, 92, 229

\BB Ford, L. A. \etal\ 1995, ApJ, 439, 307

\BB Golenetskii, S. V., \etal\ 1983, Nature, 306, 451.

\BB Graziani, C., 1992, in Gamma-Ray Bursts, ed. C. Ho, R. I. Epstein \& 
E. E. Fenimore, (Cambridge: Cambridge University Press), 407.

\BB Hurley, K. 1989, in NATO ASI Series C, Vol. 270, Cosmic Gamma Rays,
Neutrinos, and Related Astrophysics, ed. M. M. Shapiro \& J. P. Wefel 
(Dordrecht: Kluwer), 337

\BB Graziani, C., 1993, in AIP 
Conference Proceeding 280, Compton Gamma-Ray Observatory, ed. M. Friedlander, 
N. Gehrels, \& D. Macomb, (New York:AIP), 897.

\BB Kargatis, V. E., Liang, E. P., Hurley, K. C., Barat, C., Eveno, E., 
and Niel, M., 1994, ApJ, 422, 260.

\BB Katoh, M., Murakami, T., Nishimura, J., Yamagami, T., Tanaka, Y., and 
Tsunemi, H., 1984, in {\it High Energy Transients in Astrophysics}, AIP 
Conference Proc. No 115, ed. S. E. Woosley (AIP New York), 390.

\BB
Katz, J.~I., 1994, ApJ, 432, L107

\BB Laros, J. G., Evans, W. D., Fenimore, E. E., Klebesadel, R. W., 
Shulman, S., and Fritz, G., 1984, ApJ, 286, 681

\BB Lund, N., 1992, in Gamma-Ray Bursts: Observations, Analyses and 
Theories. proc. of Los Alamos Workshop, ed. C. Ho, R. Epstein, and E. Fenimore,
pg. 188.

\BB Matz, S. M., Forrest, D. J., Vestrand, W. T., Chupp, E. L., Share, G.
H., \& Rieger, E., 1985, ApJ, 288, L37.

\BB Mazets, E. P., \etal\ 1981, {\it Astrophys. Space Sci.}, { 80},
3.

\BB Mazets, E. P., Golenetskii, S. V., Ilyinskii, V. N., Aptekar, R.
L., 1982, {\it Astrophys. Space Sci.}, { 82}, 261.

\BB Metzger, A. E., Parker, R. H., Gilman, D., Peterson, L. E., \&
Trombka, J. I., 1974, ApJ, 194, L19.

\BB Metgzer, M. et al.,  1997, Nature, 387, 878

\BB Murakami, T., Ogasaka, Y., Yoshida, A, \& Fenimore, E. E., 1992a,
in AIP Conference Proceedings 265, Gamma-Ray Bursts, ed. W. S. Paciesas \& G. 
J. Fishman (New York: AIP), 28.

\BB Murakami, T., Inoue, H., van Paradijs, J., Fenimore, E., \& Yoshida,
A., 1992b, in Gamma-Ray Bursts, ed. C. Ho, R. I. Epstein \& E. E. Fenimore,
(Cambridge: Cambridge University Press), 239.

\BB Murakami, T. {\it et al.}, 1988, Nature, 335, 234.

\BB Murakami, T. {\it et al.}, 1989, {\it Publ. Astron. Soc. Jap.},
{ 41}, 405.

\BB Murakami, T. {\it et al.}, 1991, Nature, 350, 592.

\BB Norris, J. P., \etal\ 1986, ApJ, 301, 213.

\BB Ogasaka, Y., Murakami, T., Nishimura, J., Yoshida, A. \& Fenimore,
E. E., 1991, ApJ, 383, L61.

\BB Owens, A. \etal\ 1994, in Proceedings of 1993 BATSE Gamma-Ray Burst
Workshop, ed. G. J. Fishman, J. J. Brainerd, 
and K. Hurley, (New York: AIP), 665

\BB Piro, L, et al., A\&A, in press.

\BB
Preece, R., et al., 1996, ApJ, 473, 310

\BB Ricker, G. R. {\it et al.}, 1992, in, Gamma Ray Bursts:
Observations, Analyses, and Theories, {\it Proceedings of the Los Alamos 
Workshop on Gamma-Ray Bursts}, (Cambridge), ed: Ho, C., Epstein, R. I., and 
Fenimore, E. E., p. 288

\BB Schaefer, B. E., \etal\ 1994, ApJS, 92, 285

\BB Teegarden, B. J., \& Cline, T.L., 1980, ApJ, 236, L67.

\BB Wang, J. C. L., \etal\ 1989, Phys. Rev. Lett., 63, 1550.

\BB Wheaton, W. A., \etal\ 1973, ApJ, 185, L57.

\BB Yoshida, A., Murakami, T., Nishimura, J., Kondo, I. \& Fenimore, E.
E., 1992, in Gamma-Ray Bursts, ed. C. Ho, R. I. Epstein \& E. E. Fenimore,
(Cambridge: Cambridge University Press), 399.

\BB Yoshida, A., \etal\ 1989, PASJ, 41, 509.

\BB Yoshida, A., \& Murakami, T., 1994, in Proceedings of 1993 BATSE
Workshop,  ed. G. Fishman, J. Brainerd, \& K. Hurley, (New
York: AIP) pg. 333.

\vfill\eject

\vskip 10pt
\noindent{\bf Figure Captions}
\vskip 10pt

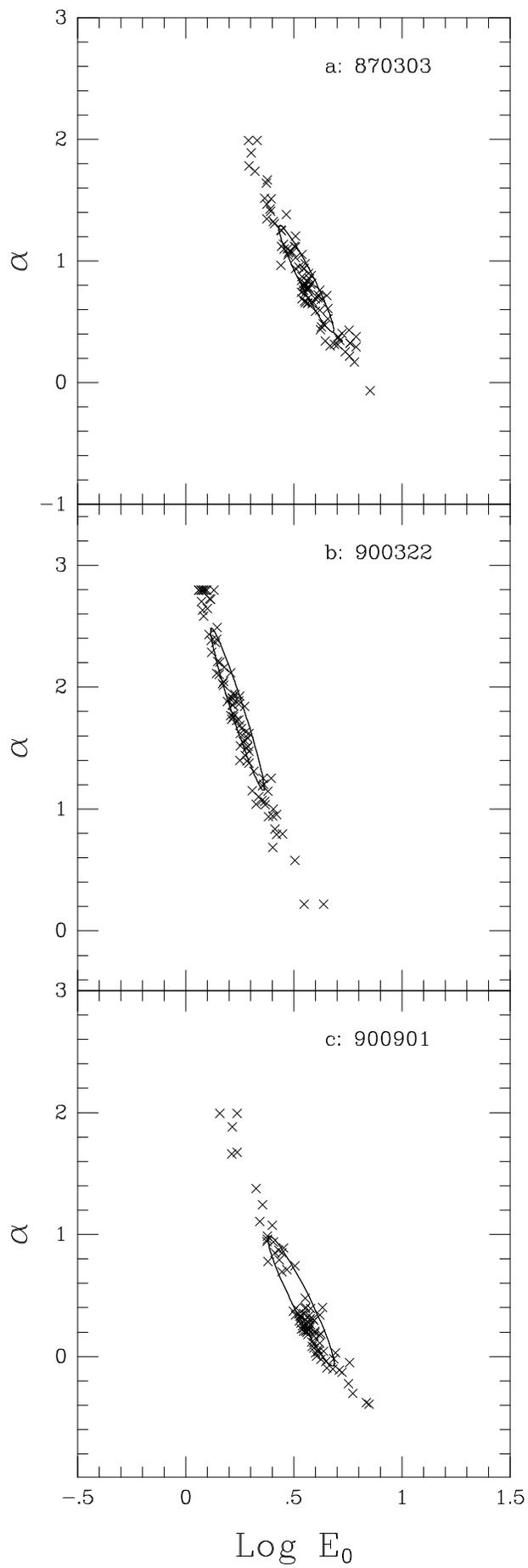
\noindent Figure 1--Simulations for three bursts to determine the effects
of the unknown incidence angle and the counting statistics. Each cross
is a simulation for a burst located at an angle
randomly selected between 0 and
60 degrees but analyzed with a response matrix corresponding to
37 degrees, the response matrix used to analyze the {\it Ginga} spectra.
One hundred simulations are shown for clarity but 10$^3$ were used.
The contour includes 68\% of the events.

\vskip 10pt

\noindent Figure 2--Best fit spectrum for each burst in our sample.
 The vertical error bars are $\pm 1\sigma$, and  the distance from
the photon spectrum to the center of the horizontal bar
represents the residual in units of $1\sigma$.
Thus, the position of the  points represent the
contribution of that data point to the $\chi^2$ of the fit (see text).
 The horizontal bars represent
the energy loss bins.  

\vskip 10pt

\noindent Figure 3a--The best-fit spectra for the 22 {\it Ginga} bursts in our
sample. Each curve has been normalized to 1.0 photons/keV at 100 keV. The 
spectra are plotted from 2.0 to 400.0 keV, 
the approximate range of sensitivity 
of the {\it Ginga} GBD.

\vskip 10pt

\noindent Figure 3b--The best-fit spectra for 54 BATSE bursts. The best-fit
parameters are taken from B93. We have plotted the spectra from 20.0 keV to
3 Mev, the approximate band pass for the BATSE spectral fits.

\vskip 10pt

\noindent Figure 4a--Simulations to determine the 68\% confidence regions
of $\alpha$ and $E_0$.  The confidence regions were found in the same
manner as in Figure 1 and include both the effects of the unknown
incidence angle and the counting statistics.
The nature of the model (eq. [1]) is such
that a range of parameters satisfying $\alpha \propto \log E_0$ can
be consistent with the data.  For two events, $E_0$ was unnecessary or
undetermined and they are shown as solid squares.

\vskip 10pt

\noindent
Figure 4b--Confidence regions reflecting the uncertainty due just to
the counting statistics.  For these simulations it was
assumed that the incidence angle was known and the same as the
analysis angle (37 degrees, for most bursts).
By comparing Figure 4a and Figure 4b, we see that 
the effects of the uncertain incidence angle is minor compared  to the effects
of the counting statistics.
The uncertain angle increases, on average,
$\log(E_0)$ by 0.09 and $\alpha$ by 0.06.

\vskip 10pt

\noindent Figure 4c--Low energy slopes and bend energies for {\it Ginga}
and BATSE events.  The open squares are 54 BATSE events from B93 and the
solid squares are the 22 {\it Ginga} events reported in this paper.
The {\it Ginga} events extend the observed range of break energies to
lower values and reveal a correlation (Pearson's $r = -0.62$)
between the low energy slopes
($\alpha$) and the bend energy ($E_0$).
For comparisons with Figure 4a, we show the confidence region for one BATSE
event (number 451).

\vskip 10pt

\noindent Figure 5a--Time history of burst 901001 
recorded in five energy bands,
two from the PC and three from the SC. The dotted vertical lines denote the 
time intervals in which spectra were computed. The dashed line is a linear 
estimate of the background derived from a fit to pre- and post-burst data. To
displace each profile vertically, 
a constant value of 600 counts has been added 
successively to each profile.

\vskip 10pt

\noindent Figure 5b--Spectra computed from the six time intervals of burst
901001 delineated in Figure 5a. 
The legend indicates the time interval for which
each spectrum was computed. The inset Figure shows the hardness-intensity
evolution of this event. 
The direction of temporal evolution is
along the solid line, from upper right to lower left.

\vskip 10pt

\noindent Figure 6a--Time history of burst 890929 recorded 
in five energy bands,
two from the PC and three from the SC. 
The dotted vertical line denote the time 
intervals in which spectra were computed.  
The dashed line is a linear estimate 
of the background derived from a fit to pre- and post-burst data. To
displace each profile vertically, 
a constant value of 600 counts has been added 
successively to each profile.

\vskip 10pt

\noindent Figure 6b--Spectra computed from the six time intervals of burst
890929 delineated in Figure 6a. 
The legend indicates the time interval for which
each spectrum was computed. The inset Figure shows the hardness-intensity
evolution of this event. 
The direction of temporal evolution is
along the solid line, from upper right to lower left.

\vskip 10pt

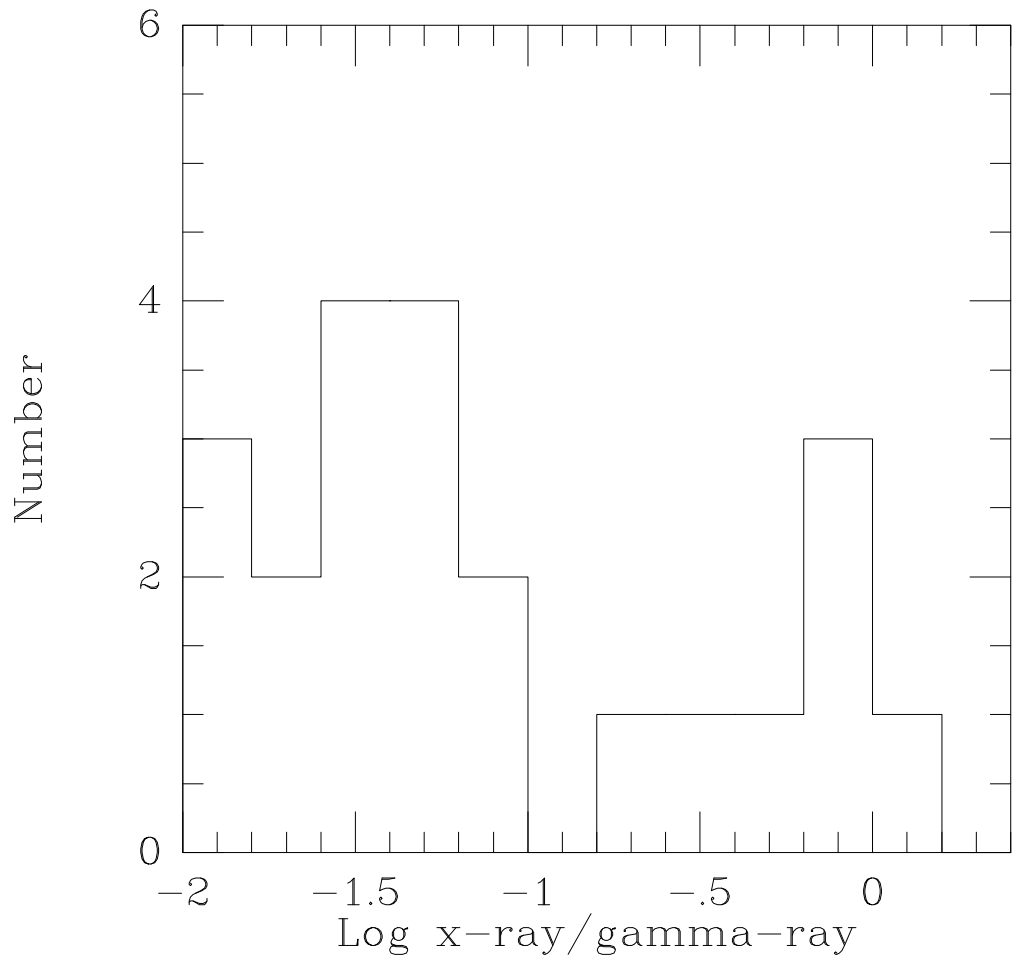
\noindent Figure 7--The distribution of the ratio of the energy emitted in
x-rays relative to that emitted in gamma-rays.  The x-ray bandpass is defined
to be from 2 to 10 kev and the gamma-ray bandpass is the BATSE range of
50 to 300 keV.  Note there are some examples of equal energy in the x-rays
and the gamma-rays.
The average ratio of the energy in the x-rays to the energy in the gamma-rays
is 24\%.

\vskip 10pt

\vfill\eject
\centerline{\bf Table 1}
\vskip 3pt
\centerline{Spectral fits to GINGA bursts} 
\vskip 10pt
\vbox{\tabskip=0pt    \offinterlineskip 

\halign to 7.5truein{\strut#\tabskip=1em plus2em&
#\hfill&
#\hfill&
#\hfill&
#\hfill&
#\hfill&
#\hfill&
#\hfill&
#\hfill\cr
& Burst$^{\rm a}$ & $\Delta T^{\rm b}$&$R_{x/g}^{\rm c}$&$E_0
^{\rm d}$ & $\alpha^{\rm e}$&$\beta^{\rm f}$&$\chi^2_r$ $^{\rm g}$ &
Ref.$^{\rm h}$ \cr
& 870303 & 5.0 & 0.182 & 3.62 & 0.74 & $-1.63\pm 0.02$
& 0.88 & 1, 3, 6, 9, 10, 11 \cr
& 870414 & 8.0 & 0.056 & 22.0 & 0.032 & $-1.80\pm 0.12$
& 1.97 & 3 \cr
& 870521 & 48.0 & 0.923 & 2.38 & 0.82 & $-2.12\pm
0.04$ & 1.47 & 3 \cr
& 870707 & 10.0 & 0.095& 53.6 & -0.67 & $-2.07\pm
0.10$ & 1.55 & 3 \cr
& 870902 & 2.5 & 0.031 & 202 & -0.76 & $-1.50\pm 0.25
$ & 1.11 & 3 \cr
& 880205 & 10.0 & 0.010 & 1997 & -0.62 & $-5.0$ &
1.49 & 1, 2, 3, 4, 8, 9, 15\cr
& 880725 & 4.0 & 0.032 & 67.7 & -0.32 & $-5.0$ & 1.03
& \cr
& 880830 & 5.0 & 0.048 & 126 & -0.75 & $-5.0$ & 1.43
& \cr
& 881009 & 16.0 & 0.303 & 4.68 &-1.46& $-1.67\pm 0.01$
& 1.36 & \cr
& 881130 & 3.0 & 0.023 & 34.0 & 0.28 & $-2.66\pm 0.61$
& 1.10 & \cr
& 890704 & 3.0 & 0.072 & 43.5 & -0.45 & $-2.05\pm 0.10$
& 1.39 & \cr
& 890929 & 5.0 & 0.027 & 118 & -0.54 & $-5.0$ & 1.26 &
8, 9 \cr
& 900126 & 6.5 & 0.670 & 1.97 & 1.63 & $-2.05\pm 0.02$
& 1.13 & 5, 7, 9, 10, 12\cr
& 900221 & 12.0 & 0.043 & 298 & -0.92 & $-5.0$ &
1.51 & \cr
& 900322a & 16.0 & 0.496 & 1.74& 1.67 & $-1.91\pm
0.05$ & 0.78 & \cr
& 900623 & 5.0 & 0.035 & 22.9 & 0.16 & $-1.63\pm 0.06$
& 1.19 & 9 \cr
& 900901 & 64.0 & 1.233 & 3.67 & 0.22 & $-2.26\pm 0.14$
& 1.71 & \cr
& 900928 & 3.5 & 0.011 & 106 & -0.21 & $-5.0$ & 1.29 &
\cr
& 901001 & 5.0 & 0.043 & 120 & -0.70 & $-5.0$ & 1.21
& \cr
& 910429 & 32.0 & 0.818 & 4.54 & -0.24 & $-2.08\pm 0.04$
& 0.91 & 13 \cr
& 910717 & 5.0 & 0.017 & 32.1 & 0.27 & $-1.69\pm 0.06$
& 1.33 & 13, 14, 16 \cr
& 910814 & 128.0 & 0.014 & 334 & -0.60 & $-5.0$
& 0.96 & 13, 14, 15, 16 \cr
}}
\vskip 5pt
\noindent $^{\rm a}$ Year/month/day of event

\noindent $^{\rm b}$ Accumulation time of spectrum in sec

\noindent $^{\rm c}$ ratio of the emitted energy in 2 to 10 kev
relative to 50 to 300 kev (see \S\ 5)

\noindent $^{\rm d}$ Break energy for Equation 1 (in keV),
uncertainties are shown in Figure 4a

\noindent $^{\rm e}$ Low energy spectral index,
uncertainties are shown in Figure 4a

\noindent $^{\rm f}$ High energy spectral index

\noindent $^{\rm g}$ Reduced $\chi^2$ for best fit spectrum (43 dof)

\noindent $^{\rm h}$ References to previous work
1: Murakami \etal\ 1988,
2: Fenimore \etal\ 1988,
3: Yoshida  \etal\ 1989,
4: Wang  \etal\ 1989,
5: Murakami \etal\ 1991,
6: Graziani \etal\ 1992,
7: Lund 1992,
8: Yoshida \etal\ 1992,
9: Murakami \etal\ 1992a,
10: Murakami \etal\ 1992b,
11: Graziani \etal\ 1993,
12: Yoshida \& Murakami: 1994,
13: Fishman \etal\ 1994,
14: Schaefer \etal\ 1994
15: Fenimore \etal\ 1993,
16: Band \etal\ 1993, Preece \etal\ 1996.
\vfill\eject

\centerline{\bf Table 2}
\vskip 3pt
\centerline{Spectral Evolution Parameters$^{\rm a}$} 
\vskip 10pt
\vbox{\tabskip=0pt    \offinterlineskip 

\halign to 7.0truein{\strut#\tabskip=1em plus2em&
#\hfill&
#\hfill&
#\hfill&
#\hfill&
#\hfill\cr
& & & 901001 & & \cr
& Time & $A$ & $E_0$ & $\alpha$ & $\beta$ \cr
& $T_1$ & 0.0398 & 122.7 & -0.810 & -5.00 \cr
& $T_2$ & 0.1056 & 121.0 & -0.702 & -5.00 \cr
& $T_3$ & 0.0295 & 101.5 & -0.976 & -5.00\cr
& $T_4$ & 0.0330 & 53.80 & -0.898 & -2.64 \cr
& $T_5$ & 0.0124 & 108.6 & -1.291 & -2.57 \cr
& $T_6$ & 0.0059 & 108.4 & -1.410 & -3.42 \cr
\cr
& & & 890929 & & \cr
& Time & $A$ & $E_0$ & $\alpha$ & $\beta$ \cr
& $T_1$ & 0.2598 & 118.2 & 0.233 & -5.00 \cr
& $T_2$ & 0.2334 & 114.9 & -0.087 & -5.00 \cr
& $T_3$ & 0.0780 & 102.1 & -0.565 & -5.00\cr
& $T_4$ & 0.0832 & 52.0 & -0.554 & -1.80 \cr
& $T_5$ & 0.1011 & 20.91 & -0.534 & -1.70 \cr
& $T_6$ & 0.5743 & 10.29 & 0.134 & -2.84 \cr
}}
\vskip 5pt

$^{\rm a}$ See \S\ 2 for a discussion of the parameters.

\bigskip


\centerline{\bf Table 3}
\vskip 3pt
\centerline{Hardness-Intensity Correlations$^{\rm a}$} 
\vskip 10pt
\vbox{\tabskip=0pt    \offinterlineskip 

\halign to 7.0truein{\strut#\tabskip=1em plus2em&
#\hfill&
#\hfill&
#\hfill&
#\hfill&
#\hfill&
#\hfill\cr
& Bursts & $a$ & $\sigma_a$ & $b$ & $\sigma_b$ & $P$ \cr
& 901001 & 0.276 & 0.035 & 3.79 & 0.208 & 0.05 \cr
& 890929 & 0.471 & 0.042 & 4.90 & 0.228 & 0.13 \cr
}}
\vskip 5pt

\noindent $^{\rm a}$ Fit of the form $\log E_p = a\log F + b$
\vfill\eject

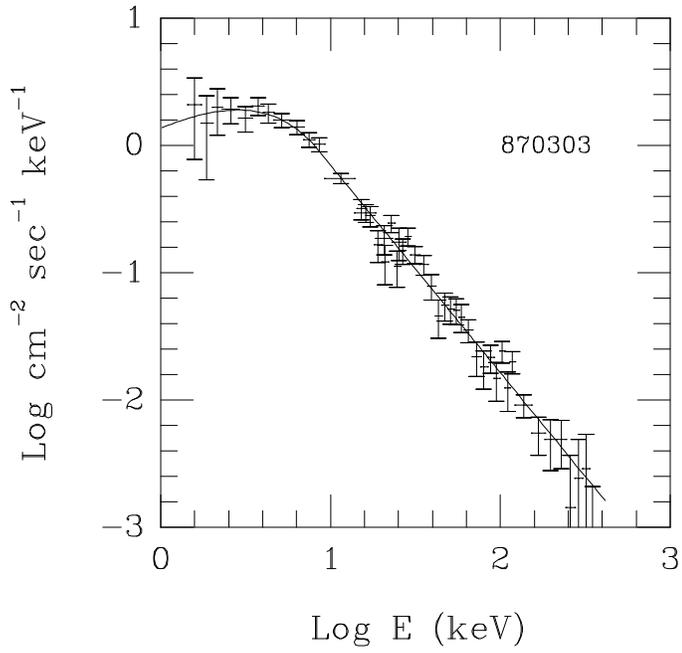
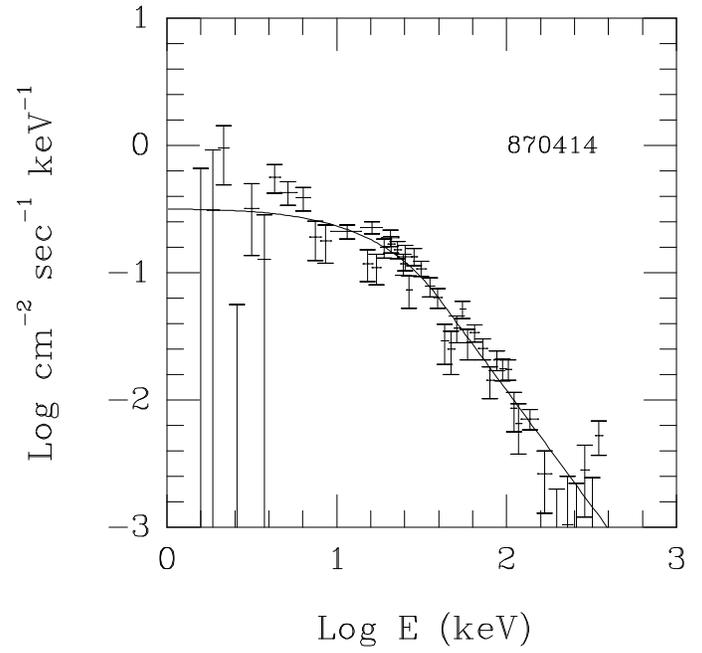
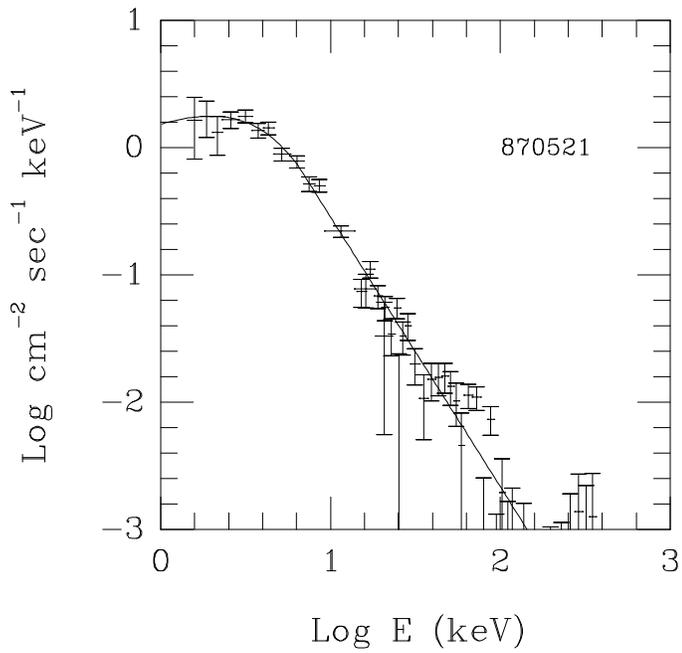
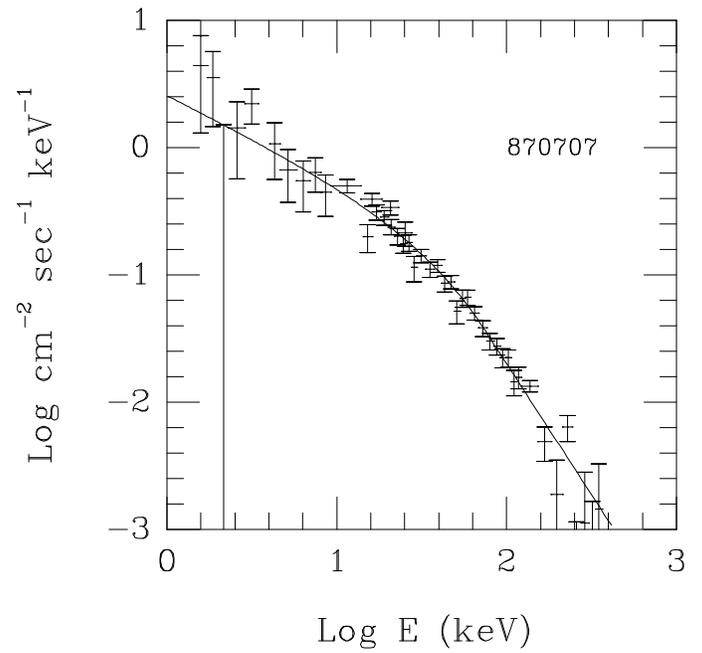
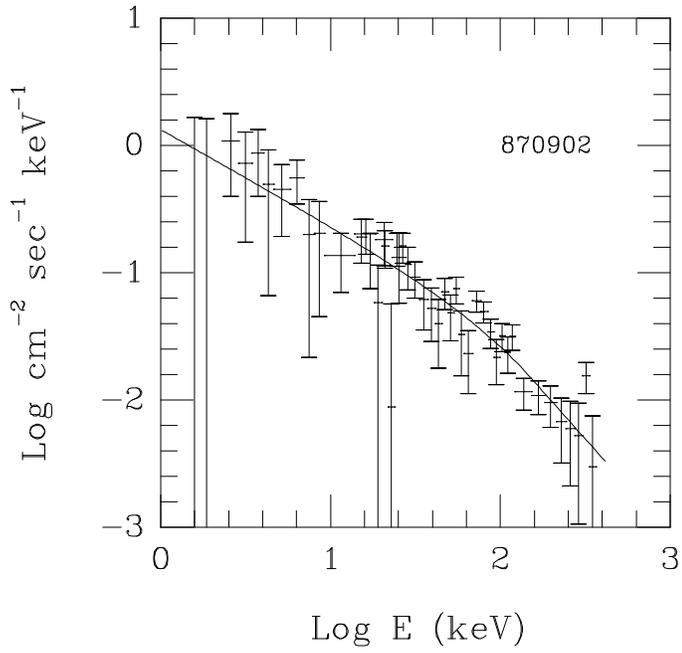
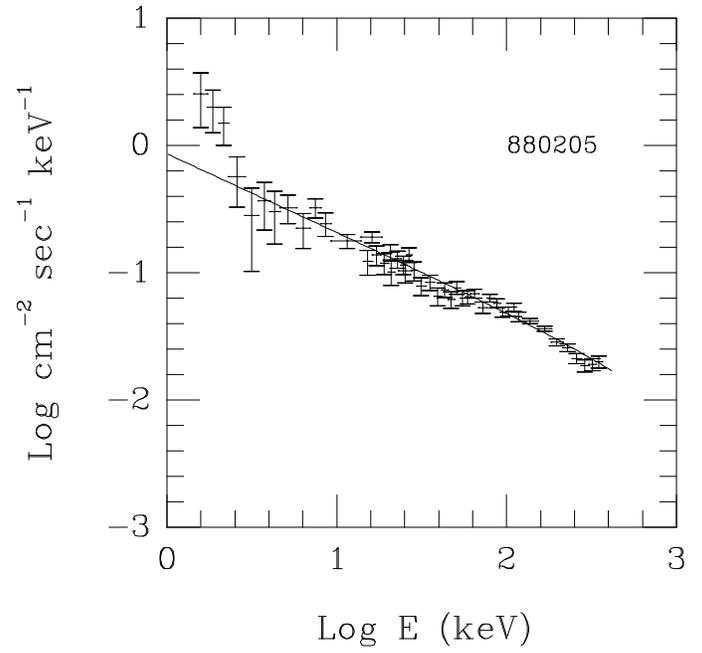
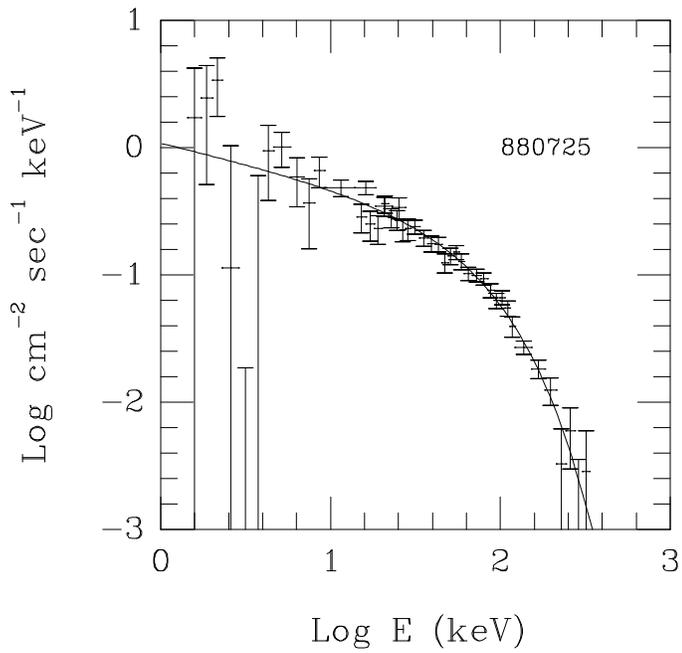
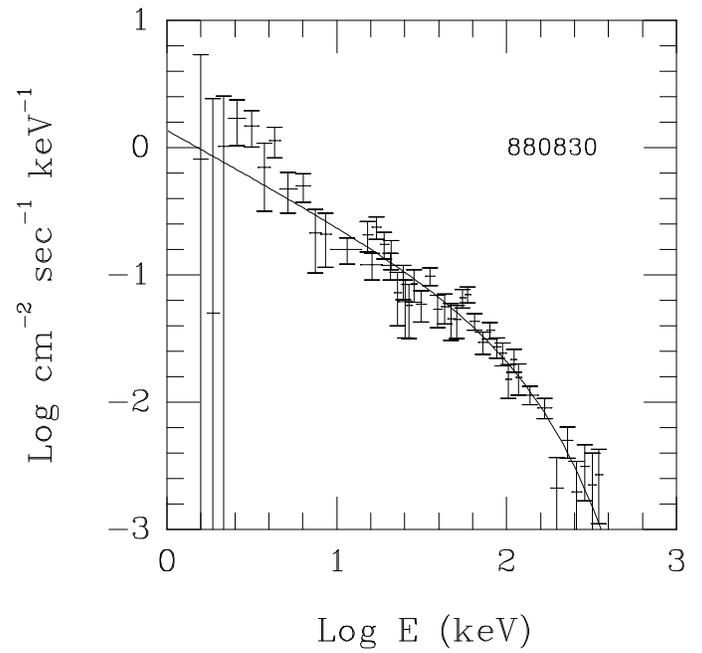
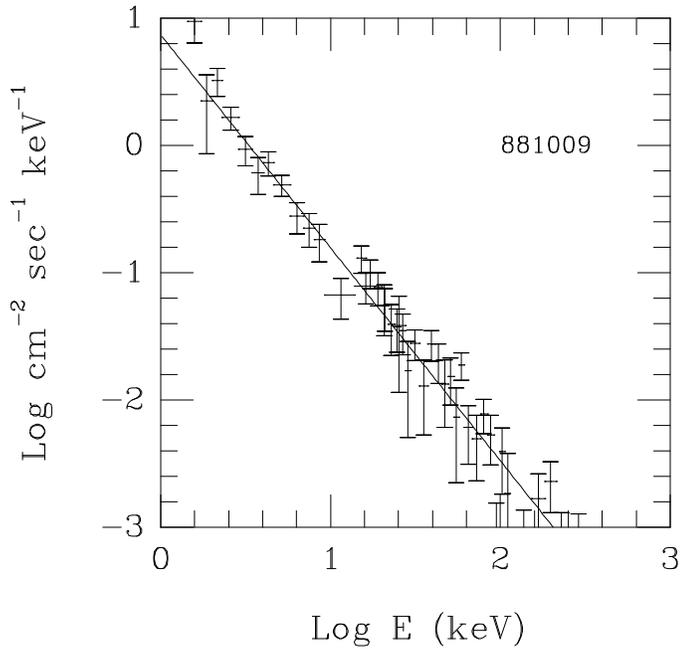
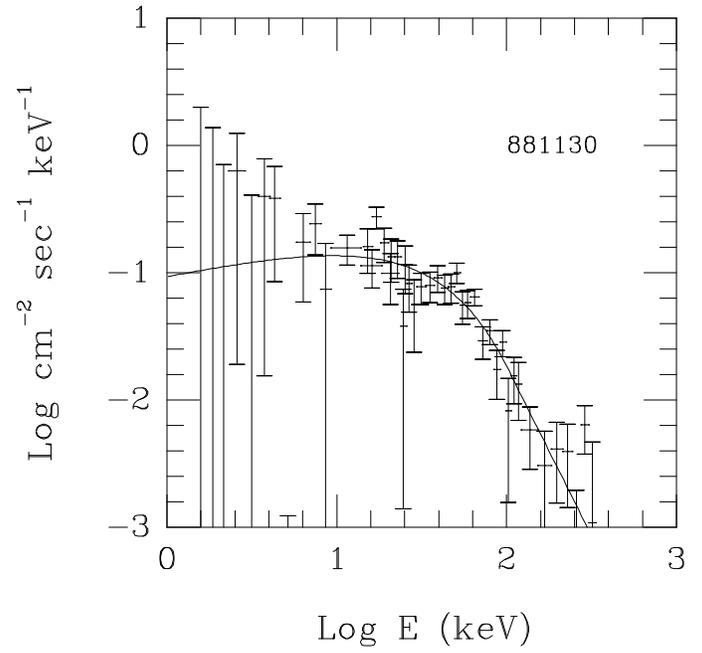
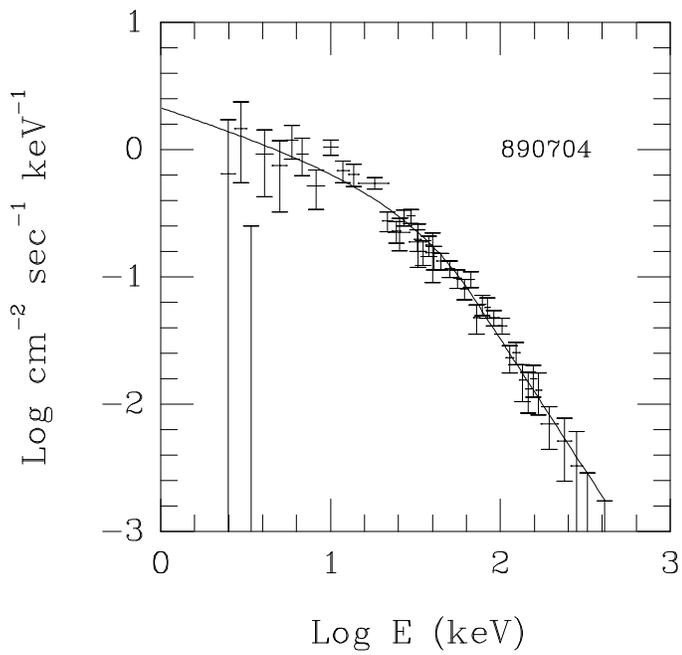
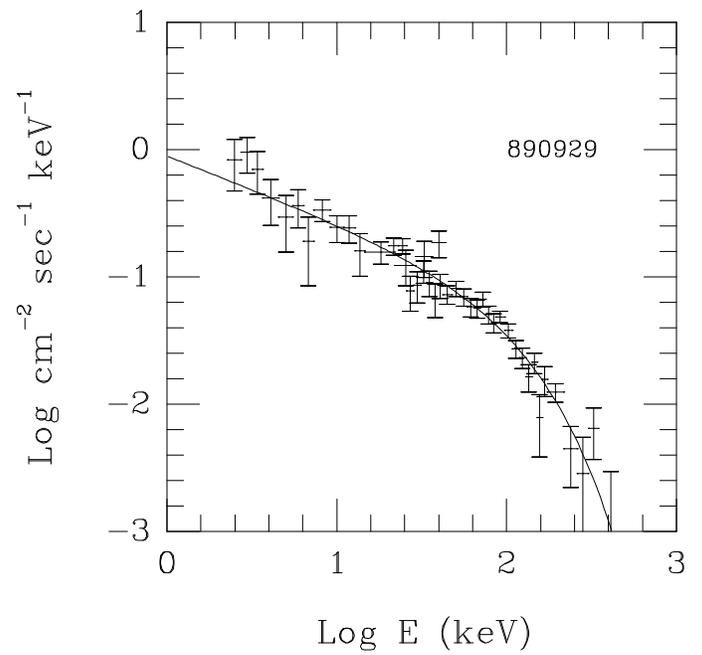
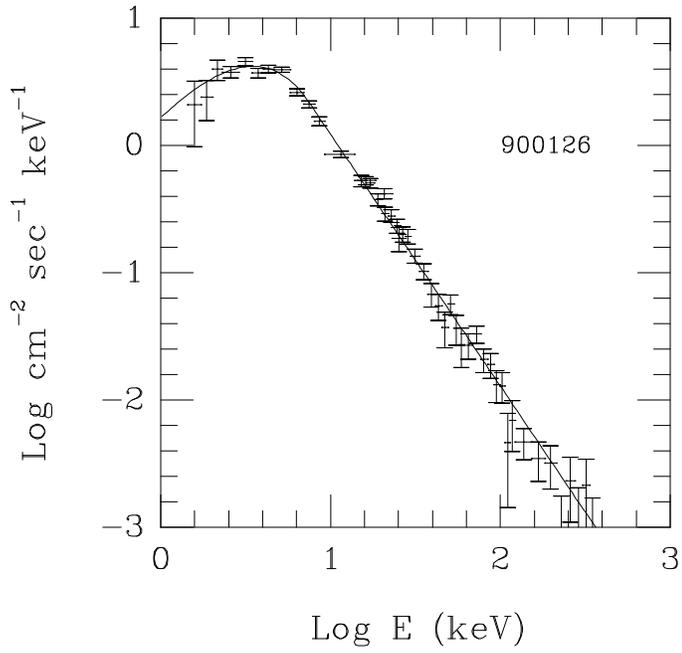
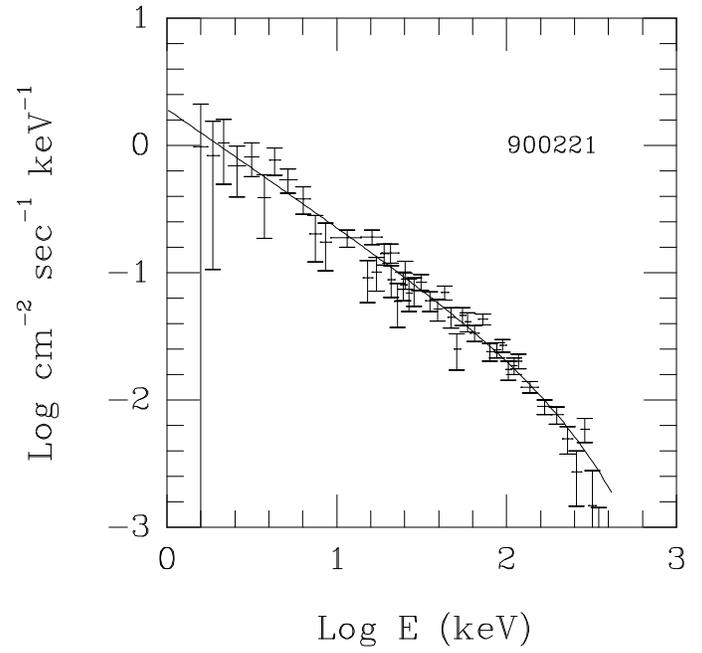
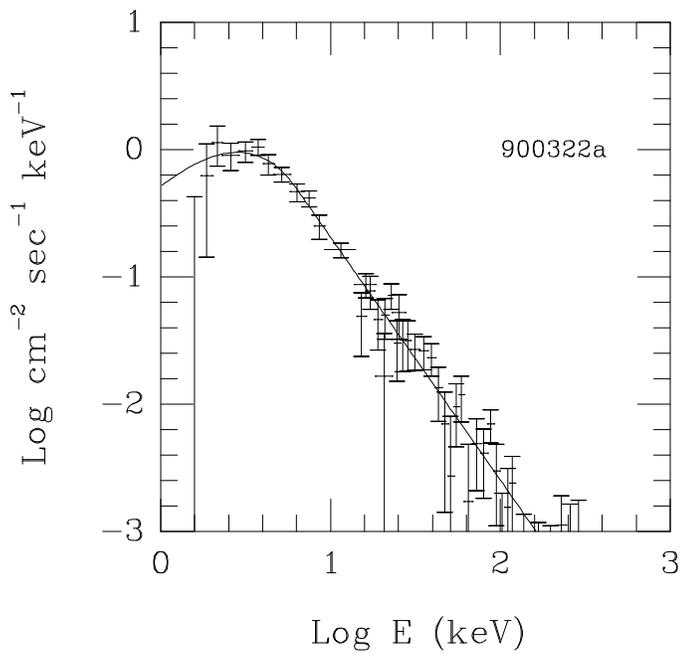
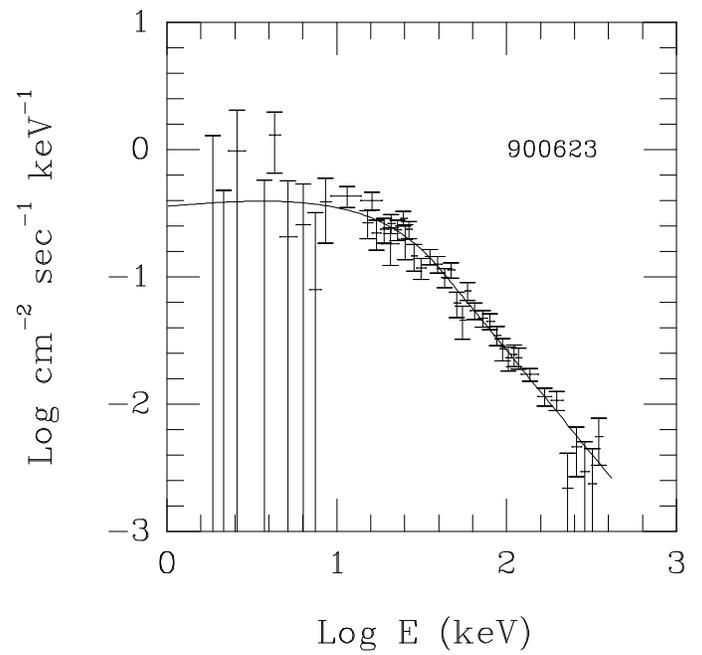
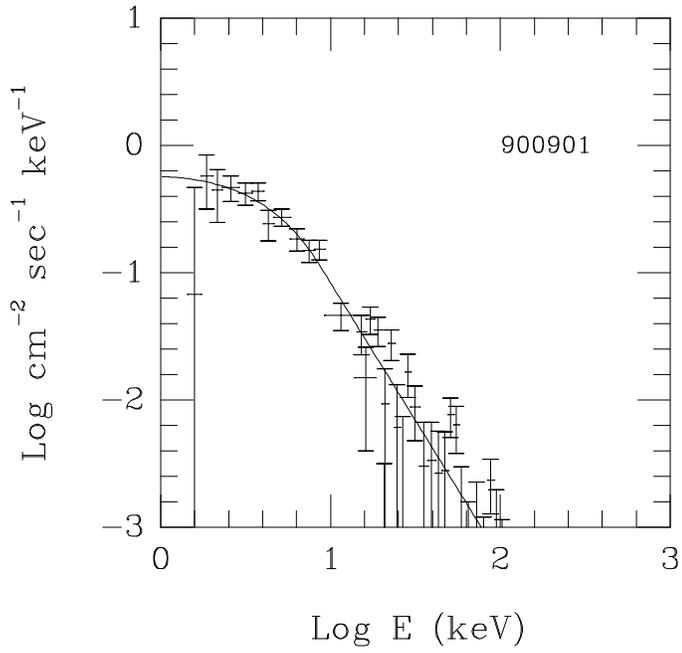
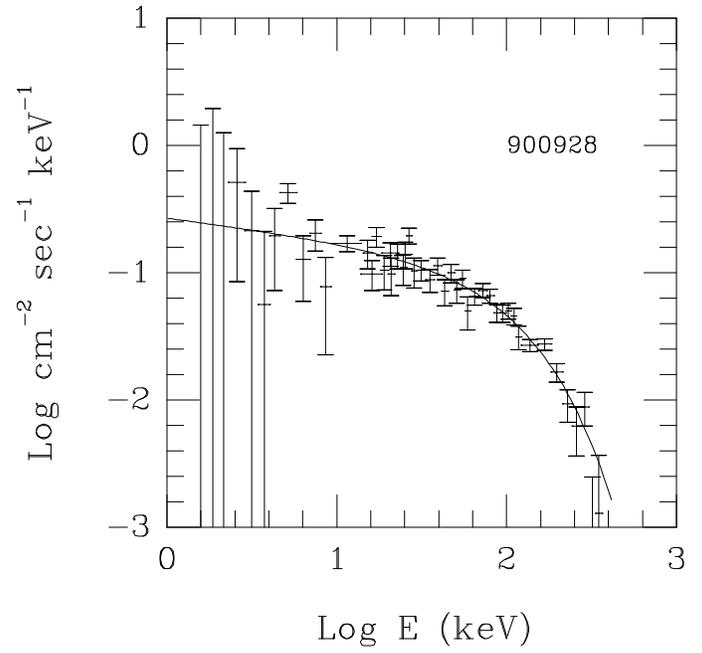
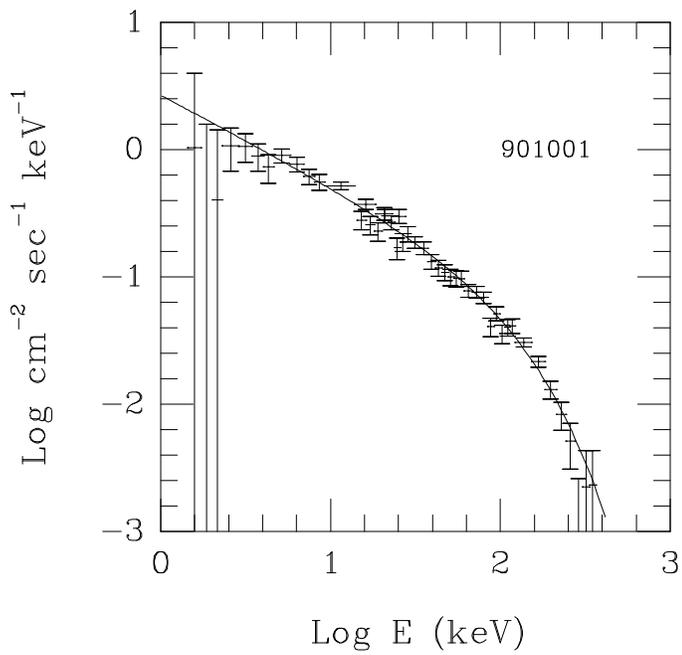
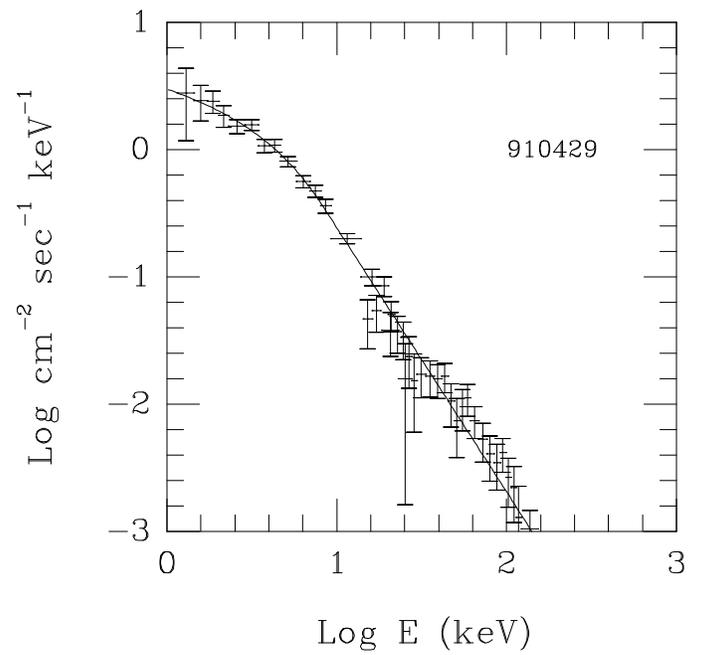
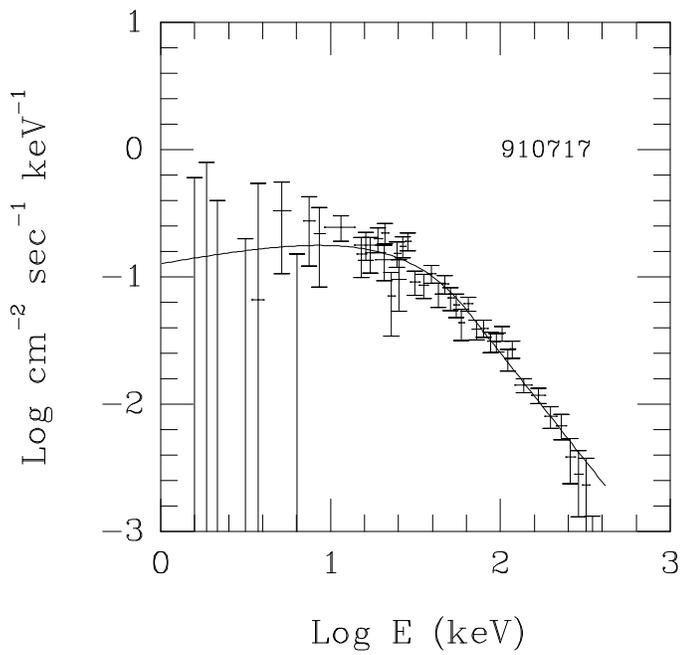
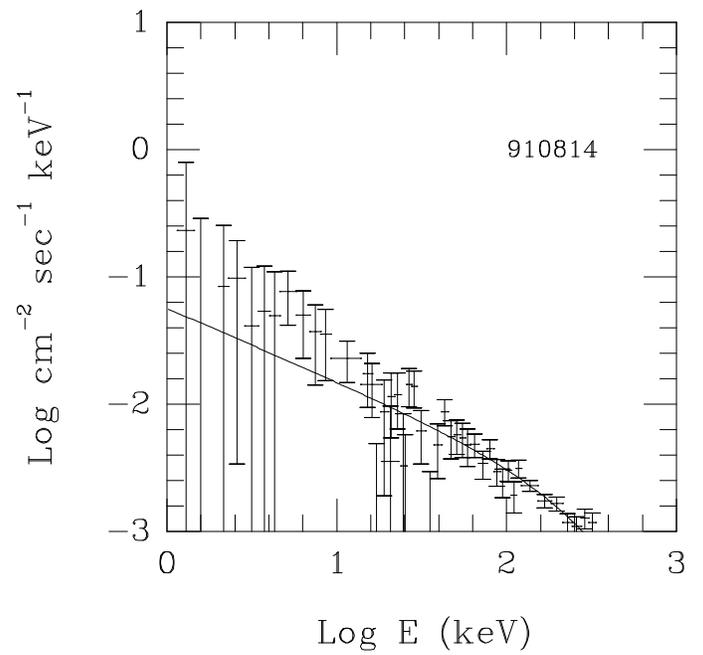
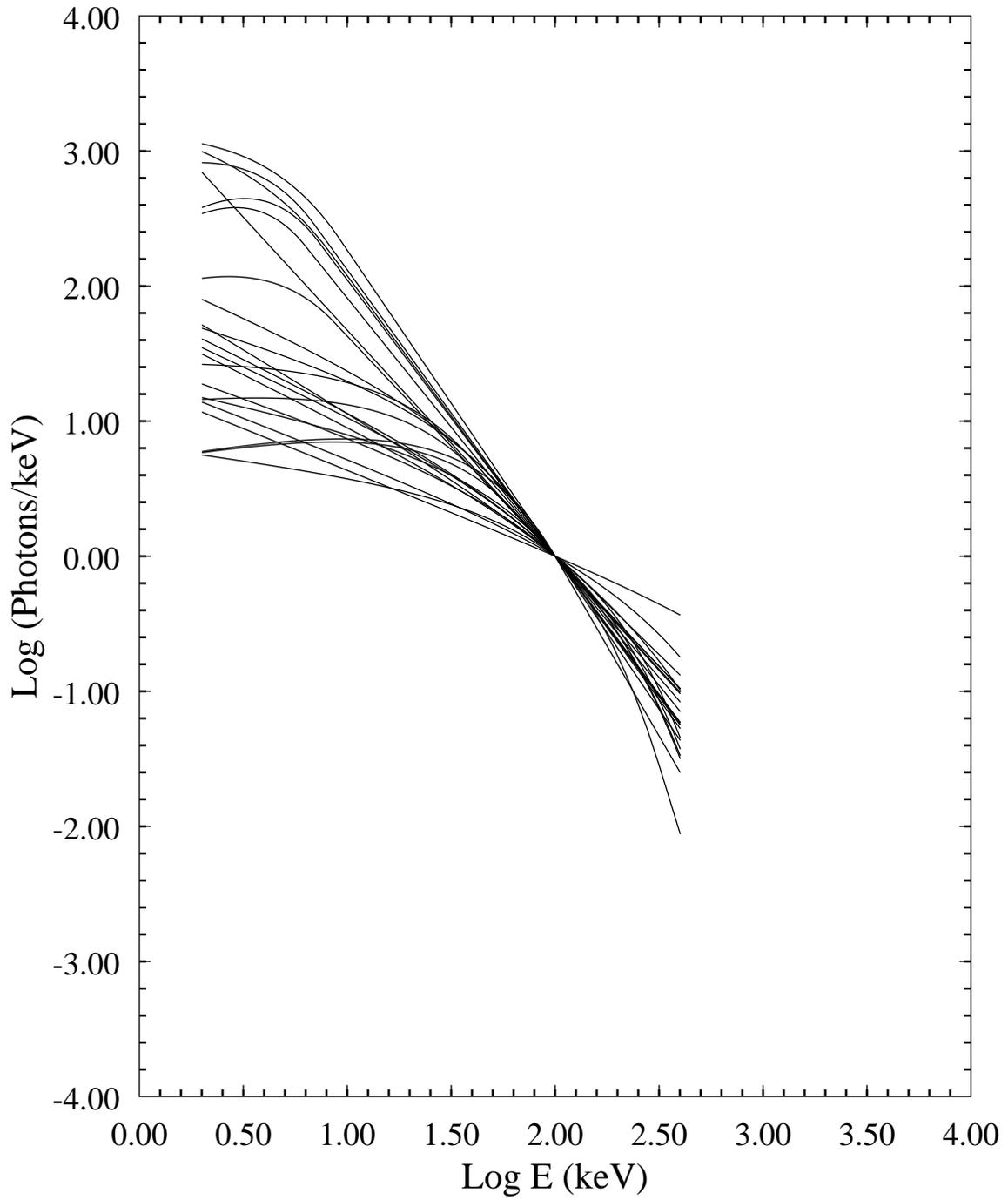
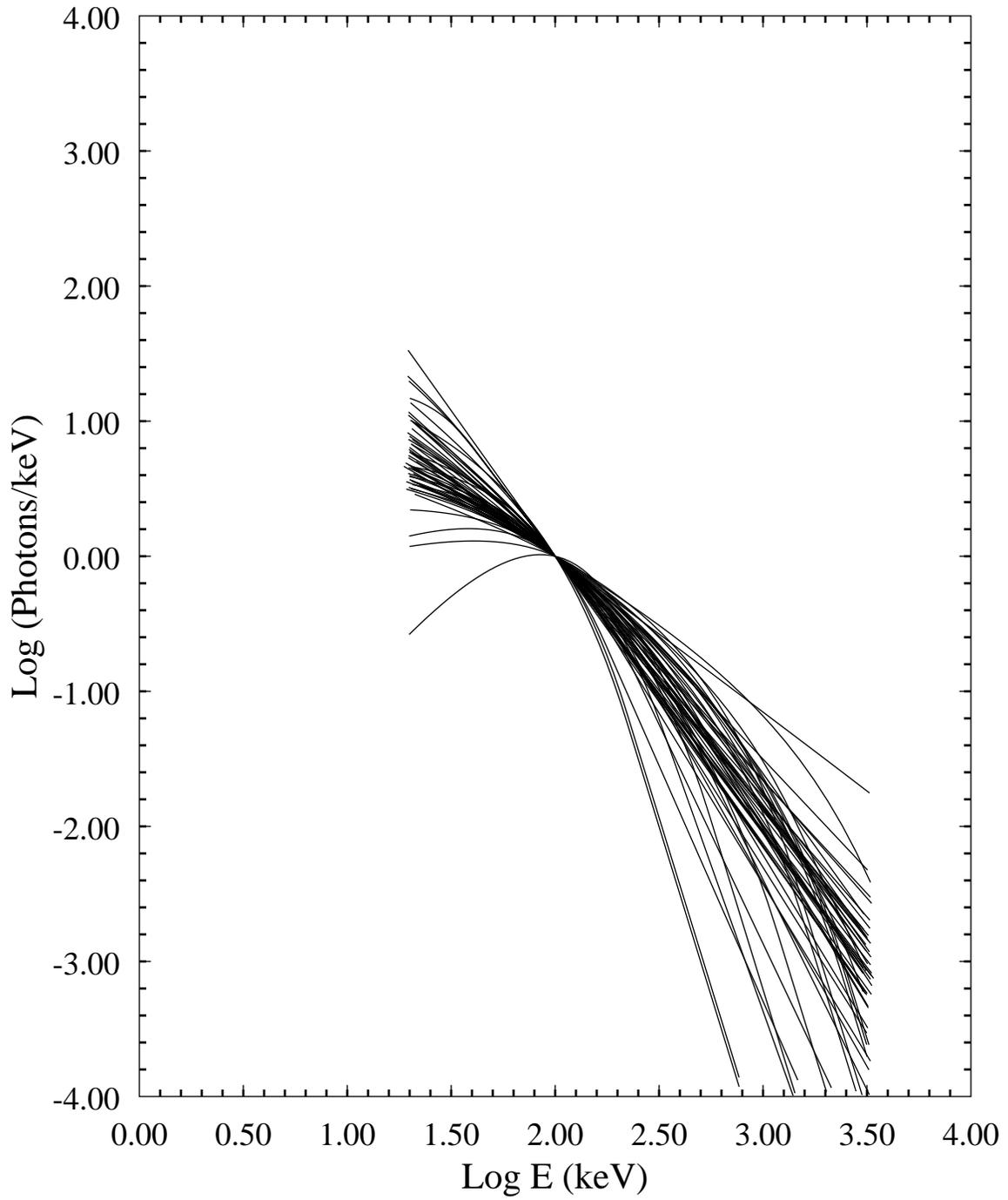
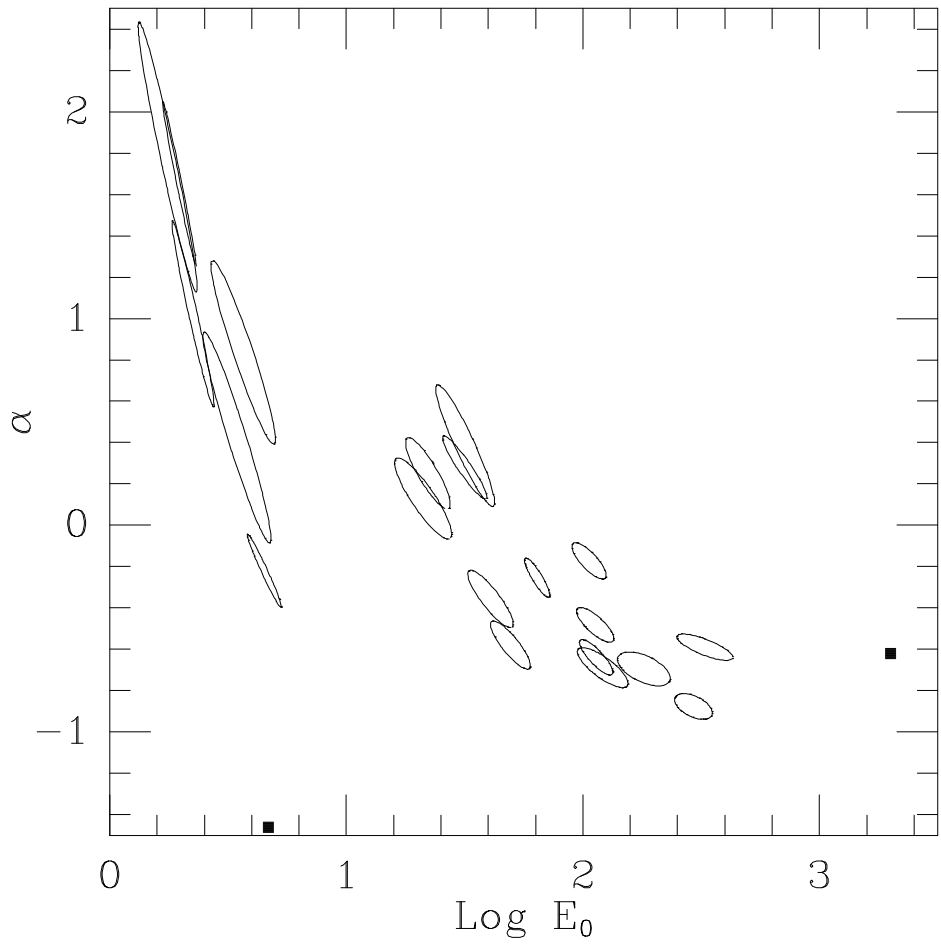
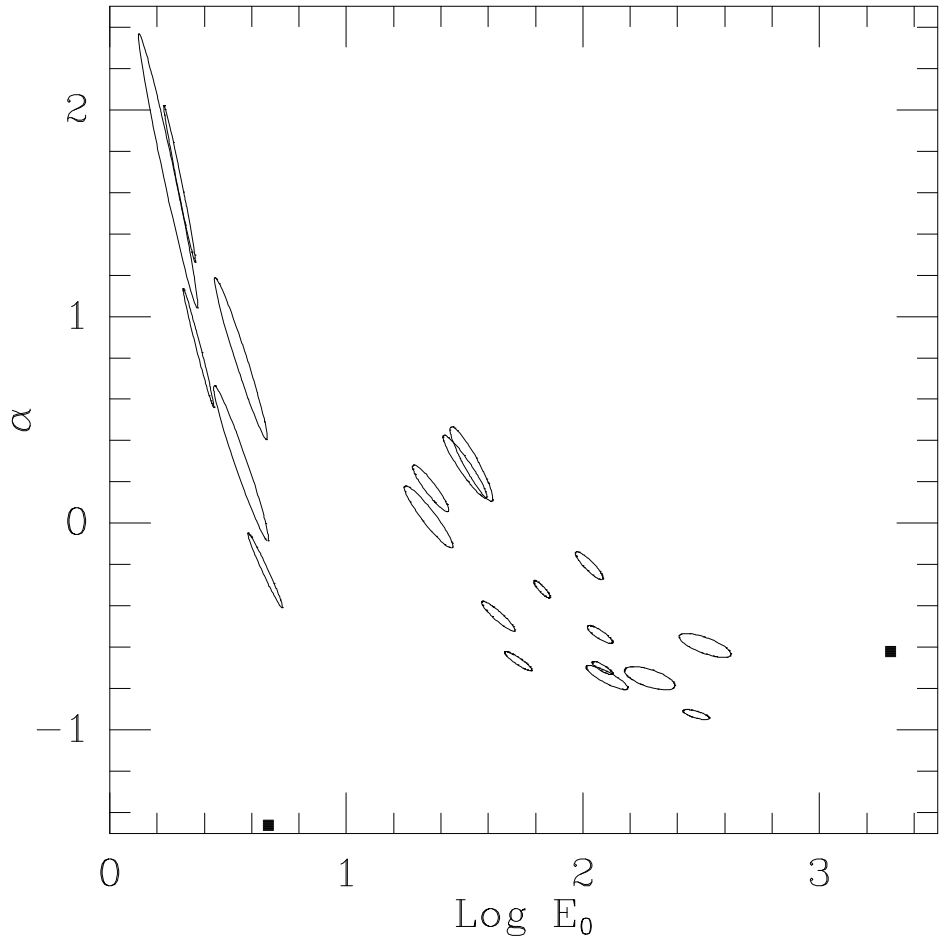
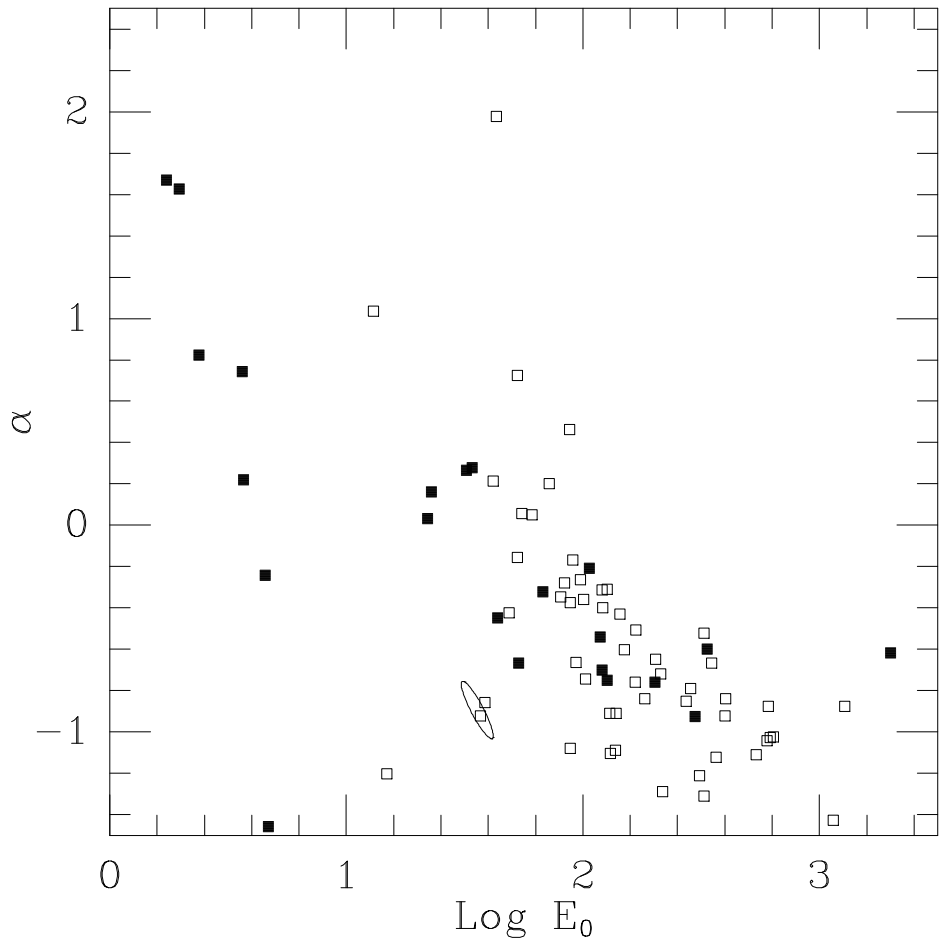
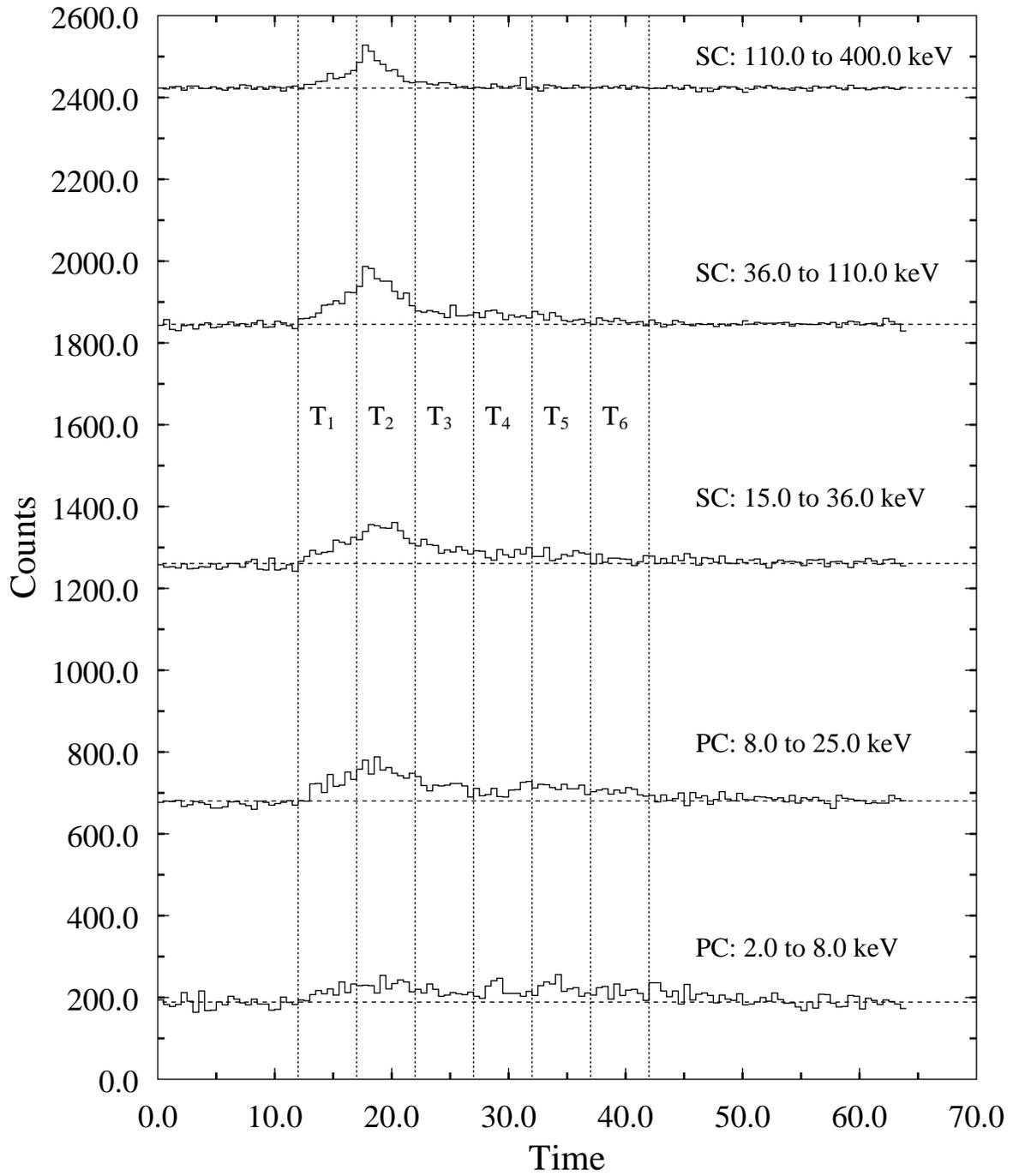
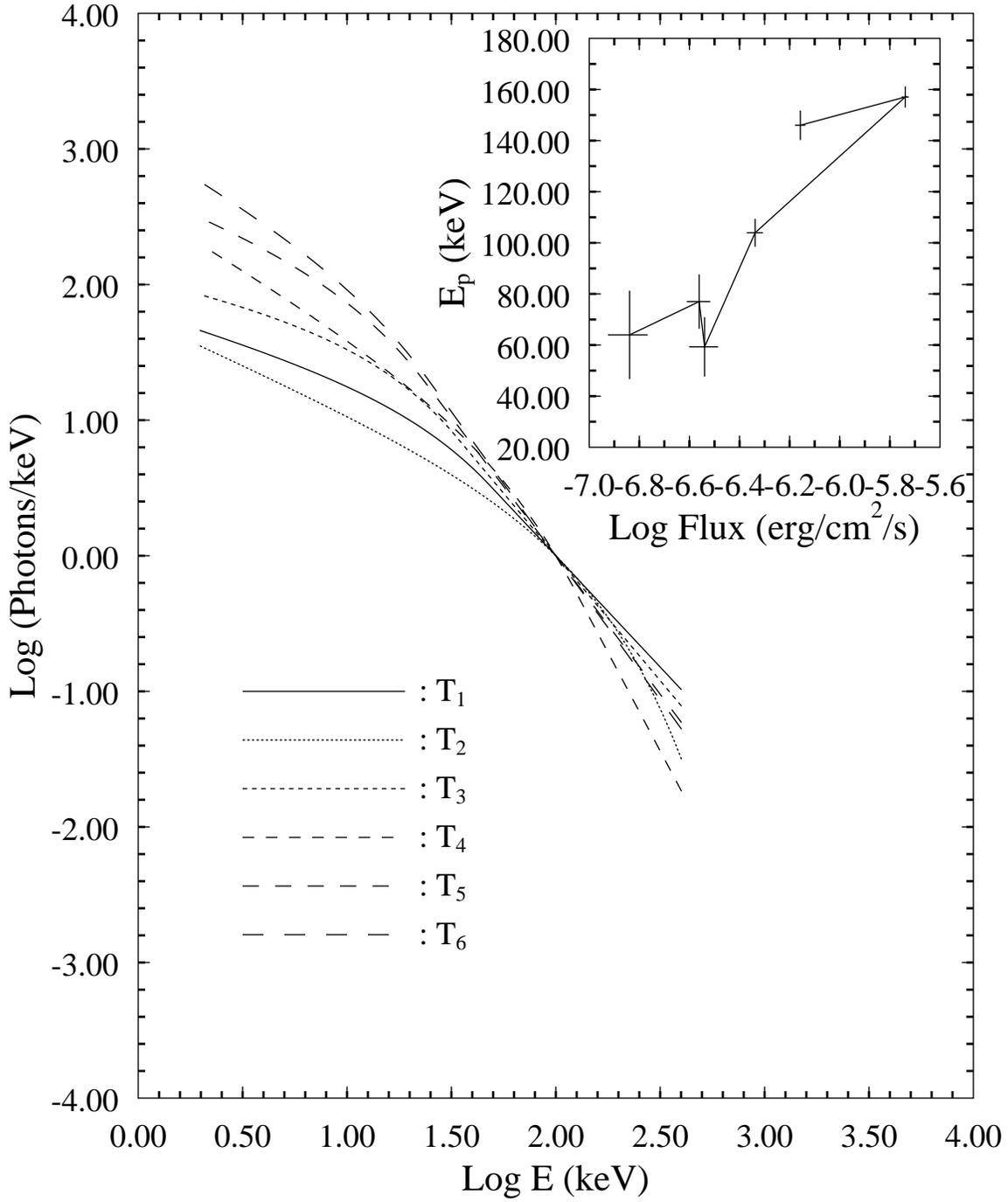
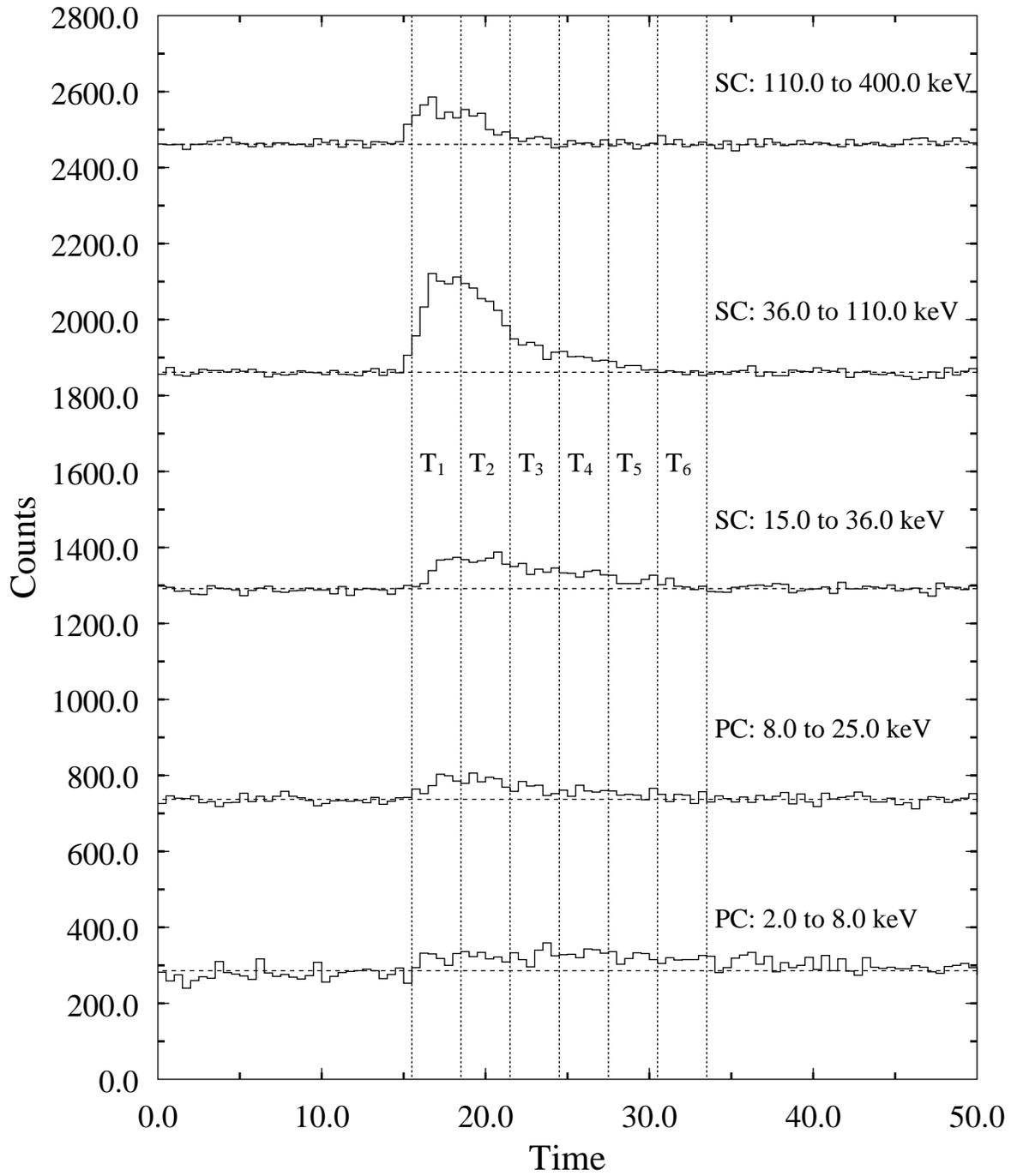
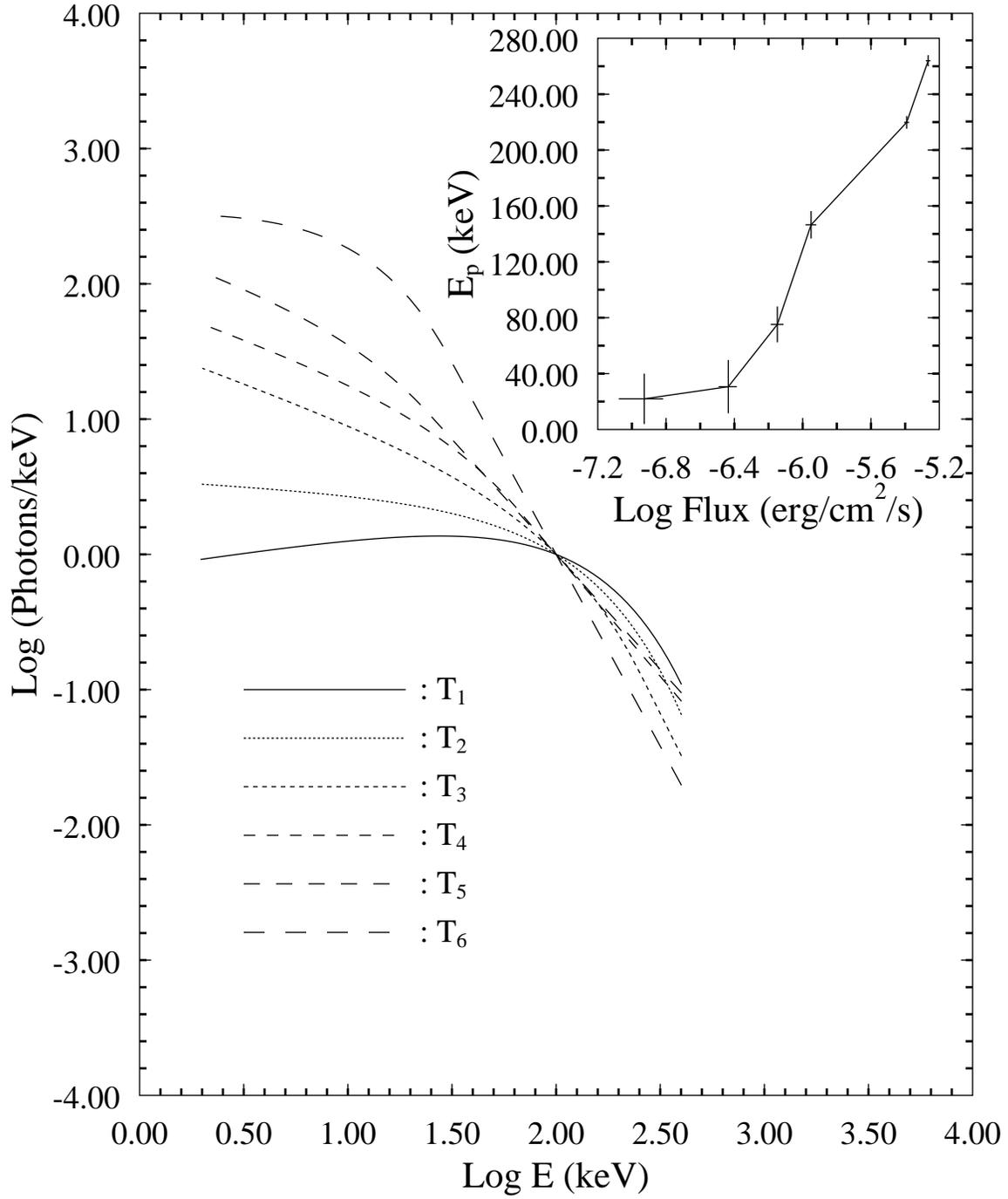
\bye